\documentclass{ws-ijmpa}
\usepackage[super,compress]{cite}
\usepackage{graphicx}
\begin{document}
\markboth{Stephan Narison}
{Improved 
$f_{D^*_{(s)}},~f_{B^*_{(s)}}$ and $f_{B_c}$ from  QCD Laplace sum rules }

\def\beq{\begin{equation}}
\def\eeq{\end{equation}}
\def\bea{\begin{eqnarray}}
\def\eea{\end{eqnarray}}
\def\bq{\begin{quote}}
\def\eq{\end{quote}}
\def\ve{\vert}
\def\nnb{\nonumber}
\def\ga{\left(}
\def\dr{\right)}
\def\aga{\left\{}
\def\adr{\right\}}
\def\rar{\rightarrow}
\def\lrar{\Longrightarrow}
\def\llar{\Longleftarrow}		
\def\nnb{\nonumber}
\def\la{\langle}
\def\ra{\rangle}
\def\nin{\noindent}
\def\ba{\vspace*{-0.2cm}\begin{array}}
\def\ea{\end{array}\vspace*{-0.2cm}}
\def\bm{\overline{m}}
\def\ind{\indexentry}
\def\c{\clubsuit}
\def\s{\spadesuit}
\def\b{$\bullet~$}
\def\als{\alpha_s}
\def\as{\ga\frac{\bar{\alpha_s}}{\pi}\dr}
\def\asr{\ga\frac{{\alpha_s}}{\pi}\dr}
\def\gg2{ \la\alpha_s G^2 \ra}
\def\gg3{g^3f_{abc}\la G^aG^bG^c \ra}
\def\ggg4{\la\als^2G^4\ra}
\def\lnu{\log{-\frac{q^2}{\nu^2}}}
\def\calD{ {\cal D} }
\def\ftilde{\tilde f}
\def\ftildeRe{ \Re e \ftilde}
\def\ftildeIm{ \Im m \ftilde}
\def\therho{\theta\rho}
\def\os{{\rm OS}}
\newcommand{\epem}{\mbox{$e^+e^-$}}
\newcommand{\dgr}{^{\rm o}}
\newcommand{\pipi}{\mbox{$\pi\pi$}}
\newcommand{\kkbar}{\mbox{$K\bar K$}}

%%%%%%%%%%%%%%%%%%%%%%%
%%%%% DEF MARINA%%%%%%%%%%%
%%%%%%%%%%%%%%%%%%%%%%%
\def\beq{\begin{equation}}
\def\enq{\end{equation}}
\def\beqa{\begin{eqnarray}}
\def\enqa{\end{eqnarray}}
\def\nnb{\nonumber}
\def\rar{\rightarrow}
\def\MeV{\nobreak\,\mbox{MeV}}
\def\GeV{\nobreak\,\mbox{GeV}}
\def\keV{\nobreak\,\mbox{keV}}
\def\fm{\nobreak\,\mbox{fm}}
\def\Tr{\mbox{ Tr }}
\def\qq{\lag\bar{q}q\rag}
\def\uu{\lag\bar{u}u\rag}
\def\dd{\lag\bar{d}d\rag}
\def\ss{\lag\bar{s}s\rag}
\def\qqs{\lag\bar{s}s\rag}
\def\mix{\lag\bar{q}g\si Gq\rag}
\def\mixs{\lag\bar{s}g\si Gs\rag}
\def\Gd{\lag G^2 \rag}
\def\gG{\lag g^2 G^2 \rag}
\def\GGG{\lag g^3G^3\rag}
\def\kab{\left[(\al+\be)m_c^2-\al\be s\right]}
\def\pli{p^\prime}
\def\ka{\kappa}
\def\lam{\lambda}
\def\La{\Lambda}
\def\gam{\gamma}
\def\Ga{\Gamma}
\def\om{\omega}
\def\rh{\rho}
\def\si{\sigma}
\def\ps{\psi}
\def\ph{\phi}
\def\de{\delta}
\def\al{\alpha}
\def\be{\beta}
\def\alma{\alpha_{max}}
\def\almi{\alpha_{min}}
\def\bemi{\beta_{min}}
\def\lb{\label}
\def\nn{\nonumber}
\def\bd{\boldmath}
\def\MS{\overline{MS}}
%%%%%%%%%%%%%%%%%%%%
\def\Fab{m_Q^2(\al+\be)-\al\be q^2}
\def\Ha{m_Q^2-\al(1-\al) q^2}
\def\Fabs{m_Q^2(\al+\be)-\al\be s}
\def\Has{m_Q^2-\al(1-\al) s}
\newcommand{\rag}{\rangle}
\newcommand{\lag}{\langle}
\newcommand{\bph}{\mbox{\bf $\phi$}}
\newcommand{\rf}{\ref}
\newcommand{\ct}{\cite}
\newcommand{\LogF}[1]{\mbox{$\:{F}_{#1} $}}
\newcommand{\LogH}[1]{\mbox{$\:{H}_{#1} $}}
\newcommand{\ImF}[1]{\mbox{$\:{\cal F}_{#1} $}}
\newcommand{\ImH}[1]{\mbox{$\:{\cal H}_{#1} $}}

%\journal{Elsevier}

%\begin{document}

%\begin{frontmatter}

%\title{$1^{--}$  heavy exotic quarkonia in QCD}
\title{ Improved 
$f_{D^*_{(s)}},~f_{B^*_{(s)}}$ and $f_{B_c}$ from  QCD Laplace sum rules 
%$^*$ \corref{cor0}
}
%\cortext[cor0]{Talk given at the 16th International Conference in QCD (QCD12), 2-6th july 2012 (Montepllier-FR). This talk  is a summary of the results in \cite{SNFB12a,SNFB12b} where more detailed discussions and more complete references can be found.}
\author{Stephan Narison}
%\email[Email address:~] {snarison@yahoo.fr}
%\thanks{}
\address{Laboratoire
Univers et Particules de Montpellier, CNRS-IN2P3, \\
Case 070, Place Eug\`ene
Bataillon, 34095 - Montpellier, France.}


\maketitle

\begin{history}
%\received{Day Month Year}
%\revised{Day Month Year}
\end{history}

\begin{abstract}
\nin
Anticipating future precise measurements of the $D$- and $B$-like  (semi-)leptonic and hadronic decays for  alternative determinations of the  CKM mixing angles,
we pursue our program on the $D$ and $B$-like mesons by improving the estimates of $f_{D^*_{(s)}}$ and $f_{B^*_{(s)}}$ (analogue to $f_\pi$) by   using the well-established (inverse) Laplace sum rules (LSR) and/or their suitable ratios less affected by the systematics, which are known to N2LO pQCD  and where the complete $d=6$ non-perturbative condensate contributions are included. The convergence of the PT series is analyzed by an estimate of the N3LO terms based on geometric growth of the coefficients. In addition to the standard LSR variable $\tau$ and the QCD continuum threshold $t_c$ stability criteria, we extract  our optimal results by also requiring  stability on the variation of the arbitrary QCD subtraction point $\mu$ which reduces the
errors in the analysis. We complete the study of the open bottom states by an estimate of $f_{B_c}$. 
Our results summarized in Tables \ref{tab:res} and \ref{tab:su3} are compared with some other recent sum rules and lattice estimates. 
 
\ccode{Pac numbers: 11.55.Hx, 12.38.Lg, 13.20-v}
\end{abstract}
%\begin{keyword}  
%QCD spectral sum rules, meson decay constants. 
%\pacs{11.55.Hx, 12.38.Lg, 13.20-v}
%\end{keyword}

%\end{frontmatter}
%%%%%%%%%%%%%%%%%%%%%%%%%%%%%
\section{Introduction}
%%%%%%%%%%%%%%%%%%%%%%%%%%%%%
\nin
The meson decay constants $f_P, f_V$ are of prime interests for understanding the realizations of chiral symmetry in QCD and for controlling the meson (semi-)leptonic decay widths, hadronic couplings and form factors. 
In addition to the well-known values of $f_\pi$=130.4(2) MeV  and $f_K$=156.1(9) MeV which control  the light flavour chiral symmetries \cite{ROSNERa,ROSNERb}, it is also desirable to extract the ones of the heavy-light charm and bottom quark systems with high-accuracy. 
This program has been initiated by the recent predictions of $f_{D_{(s)}},~f_{B_{(s)}}$ \cite{SNFB12a,SNFB12b,SNFB13} and their scalar mesons analogue \cite{SNFB3} from QCD spectral sum rules (QSSR) \cite{SVZa,SVZb}\,\footnote{For reviews where complete references can be found, see e.g: \cite{SNB1,SNB2,SNB3a,SNB3b,SNB3c,RRY,CK}.} which are improved predictions of earlier estimates \cite{SNZ,GENERALIS,SNFBa,SNFBb,SNFB4a,SNFB4b,SNFB4c,SNFB2a,SNFB2b,JAMIN3a,JAMIN3b,JAMIN3c,NEUBERTa,NEUBERTb,NEUBERTc,NEUBERTd}\, \footnote{For a recent review, see e.g. \cite{SNFBREV,SNB1}.} since the pioneering work of Novikov et al. (NSV2Z) \cite{NSV2Z} . Here, these decay constants are normalized through the matrix element:
\bea
\la 0|J_P(x)|P\ra&=& f_P M_P^2~:~J_P(x)\equiv (m_q+M_Q)\bar q(i\gamma_5)
Q~,\nnb\\
\la 0|J_V^\mu(x)|V\ra&=& f_V M_V\epsilon^\mu~:~J_V^\mu(x)\equiv \bar q\gamma_\mu Q~,
\label{eq:fp}
\label{eq:current}
\eea
where: 
$\epsilon_\mu$ is the vector polarization; $J_P(x) $ (resp $J_V^\mu(x)$) 
are the local heavy-light pseudoscalar (resp. vector) current;  $q\equiv d,c;~Q\equiv c,b;~P\equiv D,B,B_c$, $V\equiv D^*,B^*$ and where $f_P, f_V$  are related to  the leptonic widths $\Gamma [P(V)\to l^+\nu_l]$.
%\beq
%\Gamma (P^+\to l^+\nu_l)={G^2_F\over 8\pi}|V_{Qq}|^2f^2_Pm_l^2M_P\ga 1-{m_l^2\over M_P^2}\dr^2~,
%\eeq
%where $m_l$ is the lepton mass and $|V_{Qq}|$ the CKM mixing angle. 
The associated  two-point correlators are:
 \bea
\psi_{P}(q^2)&=&i\int d^4x ~e^{iq.x}\lag 0|TJ_P(x)J_P(0)^\dagger|0\rag~,\nnb\\
\Pi_V^{\mu\nu}(q^2)&=&i\int d^4x ~e^{iq.x}\lag 0
|TJ_V^{\mu}(x)J_V^{\nu}(0)^\dagger|0\rag~\nnb\\
&=&-\ga g^{\mu\nu}-{q^\mu q^\nu\over q^2}\dr\Pi^T_V(q^2)+{q_\mu q_\nu\over q^2} \Pi^L_V(q^2)~,
\lb{2po}
\eea
where one notes that $\Pi^T(q^2)$ has  more power of $q^2$ than the   transverse two-point function used in the current literature for $m_q=m_Q$ in order to avoid mass singularities at $q^2=0$ if one of the quark masses goes to zero, while $\Pi^{\mu\nu}_V(q^2)$ (vector correlator) and $\psi^{S}(q^2)$ (scalar correlator) are related each other through the Ward identities:
\beq
q_\mu q_\nu\Pi^{\mu\nu}_V(q^2)= \psi_{S}(q^2)-\psi_{S}(0)~,
 \eeq
 where to lowest order the perturbative part of $\psi_{S}(0)$ reads:
 \beq
 \psi_{S}(0)\vert_{PT}={3\over 4\pi^2}(M_Q-m_q)\ga M_Q^3Z_Q-m_q^3Z_q\dr~,
 \eeq
 with : 
 \beq
 Z_i=\ga 1- \log{M_i^2\over \mu^2}\dr\ga1+{10\over 3}a_s\dr+{2\over 3}a_s \log^2{M_i^2\over \mu^2}~,
 \eeq
 where $i\equiv Q,q$;  $\mu$ is the QCD subtraction constant and $a_s\equiv \alpha_s/\pi$ is the QCD coupling. This PT contribution which is present here has to be added to the well-known non-perturbative contribution: 
 \beq
  \psi_{S}(0)\vert_{NP}=-(M_Q-m_q)\la \bar QQ-\bar qq\ra~,
  \eeq
  for absorbing mass singularities appearing during the  evaluation of the PT two-point function, a point often bypassed in the existing literature.
Here, we extend the previous analysis of \cite{SNFB12a,SNFB12b,SNFB13,SNFB3} to the case of the $D^*_{(s)},~B^*_{(s)}$ and $B_c$ well observed mesons which have been respectively estimated earlier in \cite{SNFB88,SNB2} while $f_{B_c}$ has been also re-estimated in \cite{BAGAN,CHABABa,CHABABb}. 
The method used here will be similar to that in \cite{SNFB12a,SNFB12b,SNFB13} which are the companion  papers of this work. 
For improving the extraction of the decay constants $f_{D^*_{(s)}}$ and $f_{B^*_{(s)}}$, we shall choose to work with some suitable  ratios:
\beq
{\cal R}_{V/P}\equiv {{\cal L}_V(\tau_V)\over {\cal L}_P(\tau_P)}~,
\label{eq:ratio}
\eeq
 of the well established (inverse) Laplace sum rules\,\footnote{We use the terminology : inverse Laplace sum rule instead of Borel sum rule as it has been demonstrated in \cite{SNRAF} that its QCD radiative corrections  satisfy these properties.}$^,$\footnote{One can also work with moment sum rules like in \cite{SNFB12a,SNFB12b,SNFB13} or with $\tau$-decay like finite energy sum rules \cite{SNmassa,SNmassb} inspired from $\tau$-decay \cite{BNPa,BNPb} but these different sum rules give approximately the same results as the one from (inverse) Laplace sum rules.} :
\beq
{\cal L}_{P,V}(\tau,\mu)=\int_{(m_q+M_Q)^2}^{t_c}dt~e^{-t\tau}\frac{1}{\pi} \mbox{Im}~\big{[}\psi_P,\Pi^T_V\big{]}(t,\mu)~,
\label{eq:lsr}
\eeq
 in order to minimize  the systematics of the approach, the effects of heavy quark masses and the continuum threshold uncertainties which are one of the main sources of errors in the determinations of the decay constants. $\tau_{V,P}$ denotes the value of $\tau$ sum rule variable at which each individual sum rule is optimized (minimum or inflexion point). In general,  $\tau_{V}\not= \tau_{P}$ as we shall see later on which requires some care for a precise determination of the ratio of decay constants. This ratio of sum rule has lead to a successful prediction of the SU(3) breaking ratio of decay constants $f_{P_s}/f_P$ \cite{SNFBSU3} such that, from it, we expect to extract   precise values of the ratio $f_V/f_P$ in this paper. 
%%%%%%%%%%%%%%%%%%%%%%%%%%%%%%%%%%%
\section{QCD expression of the (inverse) Laplace Sum Rule}
%%%%%%%%%%%%%%%%%%%%%%%%%%%%%%%%%%%%
\nin
The QCD expression of the Laplace sum rule ${\cal L}_{P}(\tau,\mu)$ in the pseudoscalar channel has been already given in \cite{SNFB12a,SNFB12b} for full QCD including N2LO perturbative QCD corrections and contributions of non-perturbative condensates up to the complete $d=6$ dimension condensates and will not be repeated here\,\footnote{Note an infortunate missprint of $1/\pi$ in front of $\mbox{Im}\psi(t)$ in Ref. \cite{SNFB12a,SNFB12b}.}. 

\b To order $\alpha_s^2$, the QCD theoretical side of the LSR for the vector channel reads, in terms of the on-shell heavy quark mass $M_Q$ and for $m_d=0$:
\bea\label{eq: lsrlh}
{\cal L}_{V}(\tau)
&=& \int_{M^2_Q}^{\infty} 
{dt}~\mbox{e}^{-t\tau}\frac{1}{\pi} \mbox{Im}\Pi^T_{V}(t)\big{\vert}_{PT}
- {\la \als G^2\ra\over 12\pi}\mbox{e}^{-z}\nnb\\
%&&{\la \als G^2\ra\over 12\pi}~\mbox{e}^{-z}
&&-\Bigg{\{}\Bigg{[} 1+2a_s \Big{[}1+(1-z)\ga \ln{\nu^2\tau}+{4\over 3}\dr\Big{]}\Bigg{]}~\mbox{e}^{-z}\nnb\\
&&-2a_s\Gamma(0,z)\Bigg{\}}\ga \overline{m}_Q\over M_Q\dr^2\overline{m}_Q\la \bar dd\ra+ 
\tau ~\mbox{e}^{-z}\Bigg{\{}{z\over 4}
M_QM_0^2\la\bar dd\ra\nnb\\ 
&&+{2\over 9}\ga 1+z-\frac{z^2}{8}\dr\la\bar dj d\ra
%\frac{8\pi}{27}\rho\als \la \bar dd\ra^2\nnb\\
+\ga 1+10z-15z^2\dr {\la g^3 G^3\ra\over 8640\pi^2}\nnb\\
&&+\Big{[} 15\tilde L (8+8z-z^2)-11+124z
-200z^2
\Big{]}
{ \la j^2\ra\over 6480\pi^2}\Bigg{\}}~,
\label{eq:vec}
\eea
where:
\beq
{1\over\pi}\mbox{Im}\Pi^T_{V}(t)\big{\vert}_{PT}=\frac{1}{8\pi^2}\Big{[} t(1-x)^2(2+x)+a_s R_{1v}+a_s^2R_{2v}\Big{]}~,
\eeq
 with: $z\equiv M_Q^2\tau$; $x\equiv M^2_Q/t$; $a_s\equiv \alpha_s/\pi$; $\tilde L\equiv \ln {(\mu M_Q\tau)}+\gamma_E:  \gamma_E=0.577215...$; $\mu$ is an arbitrary subtraction point.
 
\b  $R_{1v}$ is the $\alpha_s$ corrections obtained by \cite{RRY,GENERALIS,CHETa,CHETb} and $R_{2v}$ is the $\alpha_s^2$-term obtained semi -analytically in \cite{CHETa,CHETb}.  $R_{1v}$ and  $R_{2v}$ are available as  a Mathematica package program Rvs.m. We consider as a source of errors an estimate of the N3LO assuming a geometric growth of the PT series \cite{SZ} which mimics the phenomenological $1/q^2$ dimension-two term which parametrizes the large order terms of PT series \cite{CNZa,CNZb,ZAKa,ZAKb}.   

\b The contribution up to the $d=4$ gluon condensate: 
\beq
\la\alpha_sG^2\ra\equiv \la \alpha_s G^a_{\mu\nu}G_a^{\mu\nu}\ra~,
\lb{d4g}
\eeq
of the  $d=5$ mixed condensate: 
\beq
\la\bar qGq\ra\equiv {\la\bar qg\sigma^{\mu\nu} (\lambda_a/2) G^a_{\mu\nu}q\ra}=M_0^2\la \bar qq\ra
\lb{d5mix}
\eeq
 and $d=6$ quark condensates: 
 \bea
\la\bar dj d\ra&\equiv&\la \bar d g\gamma_\mu D^\mu G_{\mu\nu}{\lambda_a\over 2} d\ra=
g^2\la \bar d \gamma_\mu {\lambda_a\over 2} d\sum_q \bar q \gamma_\mu {\lambda_a\over 2} q\ra\nnb\\
&\simeq&-{16\over 9} (\pi\alpha_s)~\rho\la \bar dd\ra^2,
\lb{eq:d6q}
\eea
after the use of the equation of motion  have been obtained originally by NSV2Z~\cite{NSV2Z}~. 

\b The contribution of the  $d=6$ gluon condensates:
\bea
\la g^3G^3\ra&\equiv& \la g^3f_{abc}G^a_{\mu\nu}G^b_{\nu\rho}G^c_{\rho\mu}\ra~,\nnb\\
 \la j^2\ra&\equiv& g^2 \la (D_\mu G^a_{\nu\mu})^2\ra
=g^4\la\ga \sum_q\bar q\gamma_\nu {\lambda^a\over 2}q\dr^2\hspace*{-0.1cm}\ra \nnb\\
&\simeq& -{64\over 3} (\pi\alpha_s)^2 \rho\la\bar dd\ra^2,
\lb{eq:d6g}
\eea
after the use of the equation of motion which are not included in the expressions given by \cite{PIVOV,LUCHA} have been deduced from the expressions given by \cite{GENERALIS} (Eqs. II.4.28 and Table II.8)
 $\rho \simeq 3-4$ measures the deviation from the vacuum saturation estimate of the $d=6$ four-quark condensates \cite{SNTAU,LNT,JAMI2a,JAMI2b,JAMI2c}. 
 
\b  One can notice that the gluon condensate $G^2$ and $G^3$ contributions flip sign from the pseudoscalar 
to the vector channel while there is an extra $m_Q M_0^2\la \bar dd\ra$ term with a positive contribution in the pseudoscalar channel. We shall see in Fig. \ref{fig:fdtau} that these different signs transform the minimum in $\tau$ for the pseudoscalar channel into an inflexion point for the vector one.  

\b  The $\alpha_s$ correction to $\la \bar dd\ra$, in the $\overline{MS}$-scheme, comes  from \cite{PIVOV}, where the running heavy quark mass  $\overline{m}_Q$ enters into this expression. 

\b Using the known relation between the running $\bar{m}_Q(\mu)$ and on-shell mass
$M_Q$ in the $\overline{MS}$-scheme to order $\alpha_s^2$ \cite{TAR,COQUEa,COQUEb,SNPOLEa,SNPOLEb,BROAD2a,BROAD2b,CHET2a,CHET2b}:
\bea
M_Q &=& \overline{m}_Q(\mu)\Big{[}
1+{4\over 3} a_s+ (16.2163 -1.0414 n_l)a_s^2\nnb\\
&&+\ln{\ga\mu\over M_Q\dr^2} \ga a_s+(8.8472 -0.3611 n_l) a_s^2\dr\nnb\\
&&+\ln^2{\ga\mu\over M_Q\dr^2} \ga 1.7917 -0.0833 n_l\dr a_s^2...\Big{]},
\label{eq:pole}
\eea
for $n_l$ light flavours, one can express all terms of the previous sum rules with the running mass $\overline{m}_Q(\mu)$.
It is clear that, for some non-perturbative terms which are known to leading order
of perturbation theory, one can use either the running or the pole 
mass. However, we shall see that this distinction does not affect, in a visible way, the present result, within the accuracy of our estimate, as the non-perturbative contributions are relatively small though vital in the analysis.
 %%%%%%%%%%%%%%%%%%%%%%%%%%%%%%%%%%%%%%%%%%
{\scriptsize
\begin{table}[hbt]
\label{tab:param}
 \tbl{QCD input parameters:
the original errors for 
$\la\alpha_s G^2\ra$, $\la g^3  G^3\ra$ and $\rho \la \bar qq\ra^2$ have been multiplied by about a factor 3 for a conservative estimate of the errors (see also the text). }  
%\tbl{
%}
%\setlength{\tabcolsep}{2pc}
    {\small
 % \begin{tabular}{lll}
 {\begin{tabular}{@{}lll@{}} \toprule
&\\
\hline
\hline
%\\
Parameters&Values& Ref.    \\
%\\
\hline
$\alpha_s(M_\tau)$& $0.325(8)$&\cite{SNTAU,BNPa,BNPb,BETHKE,PDG}\\
$\overline{m}_c(m_c)$&$1261(12)$ MeV &average \cite{SNH10a,SNH10b,SNH10c,PDG,IOFFEa,IOFFEb}\\
$\overline{m}_b(m_b)$&$4177(11)$ MeV&average \cite{SNH10a,SNH10b,SNH10c,PDG}\\
$\hat \mu_q$&$(253\pm 6)$ MeV&\cite{SNB1,SNmassa,SNmassb,SNmass98a,SNmass98b,SNLIGHT}\\
%$\la \bar dd\ra(2) $&$-(275.7\pm 6.6)^3$ MeV$^3$&\cite{SNB1,SNmass}\\
$M_0^2$&$(0.8 \pm 0.2)$ GeV$^2$&\cite{JAMI2a,JAMI2b,JAMI2c,HEIDa,HEIDb,HEIDc,SNhl}\\
$\la\alpha_s G^2\ra$& $(7\pm 3)\times 10^{-2}$ GeV$^4$&
\cite{SNTAU,LNT,SNIa,SNIb,YNDU,SNHeavy,BELLa,BELLb,BELLc,SNH10a,SNH10b,SNH10c,SNG1,SNG2,SNGH}\\
$\la g^3  G^3\ra$& $(8.2\pm 2.0)$ GeV$^2\times\la\alpha_s G^2\ra$&
\cite{SNH10a,SNH10b,SNH10c}\\
$\rho \alpha_s\la \bar qq\ra^2$&$(5.8\pm 1.8)\times 10^{-4}$ GeV$^6$&\cite{SNTAU,LNT,JAMI2a,JAMI2b,JAMI2c}\\
$\hat m_s$&$(0.114\pm0.006)$ GeV &\cite{SNB1,SNTAU9,SNmassa,SNmassb,SNmass98a,SNmass98b,SNLIGHT}\\
$\kappa\equiv \la \bar ss\ra/\la\bar dd\ra$& $(0.74^{+0.34}_{- 0.12})$&\cite{HBARYONa,HBARYONb,SNB1}\\
\hline\hline
\end{tabular}}
}
%\caption{%\scriptsize   
\end{table}
} 
%%%%%%%%%%%%%%%%%%%%%%%%%%%%%%
%\end{document}
%%%%%%%%%%%%%%%%%%%%%%%%%%%%%%%%%%%
\section{QCD input parameters}
%%%%%%%%%%%%%%%%%%%%%%%%%%%%%%%%%%%
\nin
The QCD parameters which shall appear in the following analysis will be the charm and bottom quark masses $m_{c,b}$ (we shall neglect  the light quark masses $q\equiv u,d$),
the light quark condensate $\qq$,  the gluon condensates $ \lag
\alpha_sG^2\rag$
%\equiv \la \alpha_s G^a_{\mu\nu}G_a^{\mu\nu}\ra$ 
and $ \la g^3G^3\ra$
%\equiv \la g^3f_{abc}G^a_{\mu\nu}G^b_{\nu\rho}G^c_{\rho\mu}\ra$, 
the mixed condensate $\la\bar qGq\ra$ defined in Eq. (\ref{d4g}) to Eq. (\ref{eq:d6g})
%\equiv {\la\bar qg\sigma^{\mu\nu} (\lambda_a/2) G^a_{\mu\nu}q\ra}=M_0^2\la \bar qq\ra$ 
and the four-quark 
 condensate $\rho\alpha_s\la\bar qq\ra^2$, where
 $\rho\simeq 3-4$ indicates the deviation from the four-quark vacuum 
saturation. Their values are given in Table \ref{tab:param}. 

\b We shall work with the running
light quark condensates and masses. 
%parameters estimated to order $\alpha_s^3$ \cite{SNB1,SNB2}. 
They read:
\bea
%{\bar m}_{q,Q}(\tau)&=&
%{\hat m}_{q,Q}  \ga-\beta_1a_s\dr^{-2/{
%\beta_1}}\times C(a_s)
%\nnb\\
{\la\bar qq\ra}(\tau)&=&-{\hat \mu_q^3  \ga-\beta_1a_s\dr^{2/{
\beta_1}}}/C(a_s)
\nnb\\
{\la\bar q Gq\ra}(\tau)&=&-{M_0^2{\hat \mu_q^3} \ga-\beta_1a_s\dr^{1/{3\beta_1}}}/C(a_s)~,\nnb\\
\overline{m}_s(\tau)&=&{\hat m_s\over{\ga -{\rm Log}{\sqrt{\tau}\Lambda}\dr ^{2/-\beta_1}}}C(a_s)~,
\eea
where $\beta_1=-(1/2)(11-2n_f/3)$ is the first coefficient of the $\beta$ function 
for $n_f$ flavours; $a_s\equiv \alpha_s(\tau)/\pi$; 
%${\hat m}_{q,Q}$ is the RGI quark mass, 
$\hat\mu_q$ is the spontaneous RGI light quark condensate \cite{FNR}. The QCD correction factor $C(a_s)$ in the previous expressions is numerically \cite{RUNDEC}:
\bea
C(a_s)&=& 1+0.8951a_s+1.3715a_s^2 +...~~{\rm :}~~ n_f=3~,\nnb\\
&=&1+1.1755a_s+1.5008a_s^2 +...~~{\rm :}~~ n_f=5~,
\eea
which shows a good convergence. 
%For the extraction of $f_{D_{(s)}, B_{(s)}}$, 
We shall use:
\beq   
\alpha_s(M_\tau)=0.325(8) \lrar  \alpha_s(M_Z)=0.1192(10)
\label{eq:alphas}
\eeq
from $\tau$-decays \cite{SNTAU,BNPa,BNPb}, which agree perfectly with the world average 2012 \cite{BETHKE,PDG}: 
\beq
\alpha_s(M_Z)=0.1184(7)~. 
\eeq
%We shall also use the value of the running strange quark mass obtained in \cite{SNmass}\,\footnote{This value agrees and improves previous sum rules results \cite{SNmass2}.} given in Table~\ref{tab:param}. 
The value of the running $\la \bar qq\ra$ condensate is deduced from  the well-known GMOR relation: 
\beq
(m_u+m_d)\la \bar uu+\bar dd\ra=-m_\pi^2f_\pi^2~,
\eeq
where $f_\pi=130.4(2)$ MeV \cite{ROSNERa,ROSNERb} and the value of $(\overline{m}_u+\overline{m}_d)(2)=(7.9\pm 0.6)$ MeV obtained in  \cite{SNmassa,SNmassb} which agrees with the PDG  in \cite{PDG}  and lattice averages in \cite{LATT13}. Then, we deduce the RGI light quark spontaneous mass $\hat\mu_q$ given  in Table~\ref{tab:param}. 
%\beq
%\hat m_s=128(7) ~{\rm MeV},~~~~~~~~~~~~~
%\hat\mu_q=263(7)~{\rm MeV}.
%\label{eq:lightmass}
%\eeq

\b For the heavy quarks, we shall use the running mass and the corresponding value of $\alpha_s$ evaluated at the scale $\mu$. These sets of correlated parameters are given in Table \ref{tab:alfa} for different values of $\mu$ and for a given number of flavours $n_f$.

\b For the $\la \alpha_s G^2\ra$ condensate, we have the enlarged the original error by a factor about 3 in order to have
a conservative result and to recover the original SVZ estimate and the alternative extraction in \cite{IOFFEa,IOFFEb} from charmonium sum rules which we consider as the most reliable channel for extracting phenomenologically this condensate.  However, a direct comparison of this  range of values obtained within short QCD series (few terms) with the one from lattice calculations \cite{BALIa} obtained within a long QCD series remains to be clarified\cite{BALIb}. 

\b To be conservative, we have also enlarged the original error on the value of the $SU(3)$ breaking condensate $\kappa\equiv \la\bar ss\ra/\la\bar dd\ra$ given in \cite{HBARYONa,HBARYONb}  to recover the central value 1.08 from lattice calculation \cite{UKQCD}.

\b Some other estimates of the gluon and four-quark condensates using $\tau$-decay and $e^+e^-\to I=1$ hadrons data can be found in \cite{DAVIER,BOITOb,FESRa,FESRb}. Due to the large uncertainties induced by the different resummations of the QCD series and by the less-controlled effects of some eventual duality violation, we do not consider explicitly these values in the following analysis. However, we shall see later on that the effects of the gluon and four-quark condensates on the values of the decay constants are almost negligible though they play an important r\^ole in the stability analysis. 

%%%%%%%%%%%%%%%%%%%%%%%%%%%%%%%%%%%%%%%%%%
{\scriptsize
\begin{table}[hbt]
 \tbl{%\scriptsize    
$\alpha_s(\mu)$ and correlated values of $\overline{m}_Q(\mu)$ used in the analysis for different values of the subtraction scale $\mu$. The error in $\overline{m}_Q(\mu)$ has been induced by the one of $\alpha_s(\mu)$ to which one has added the error on their determination given in Table\,\ref{tab:param}. }
%\setlength{\tabcolsep}{1.2pc}
%The original errors have been multiplied by 2 for a conservative estimate of the errors.    
    {\small
%\begin{tabular}{llll}
{\begin{tabular}{@{}llll@{}} \toprule
&\\
\hline
\hline
%\\
Input for $f_{D^*_{(s)}}$ : $n_f=4$\\
\hline
$\mu$[GeV]&$\alpha_s(\mu)$&& $\overline{m}_c(\mu)$[GeV]\\
%\hline
% \multicolumn{2}{l}{}&&\\
% \hline
1&0.4896(223)&&1.422(12)\\
Input: $\overline{m}_c(m_c)$&0.4084(144)&&1.26\\
1.4&0.3804(125)&&1.206(2)\\
1.45&0.3725(116)&&1.191(4)\\
1.5&0.3649(110)&&1.176(5)\\
1.55&0.3579(105)&&1.162(6)\\
1.6&0.3513(101)&&1.148(5)\\
2&0.3120(77)&&1.069(9)\\
2.5&0.2812(61)&&1.005(10)\\
%3&0.2606(51)&&0.961(10)\\
%3.5&0.2455(45)&&0.929(11)\\
\hline
Input for $f_{B^*_{(s)}}$: $n_f=5$\\
\hline
$\mu$[GeV]&$\alpha_s(\mu)$& $\overline{m}_b(\mu)$[GeV]&\\
3&0.2590(26)&4.474(4)&\\
Input: $\overline{m}_b(m_b)$&0.2320(20)&4.177&\\
4.5&0.2267(2)&4.119(1)&\\
5&0.2197(18)&4.040(1)&\\
6&0.2085(16)&3.914(2)&\\
7&0.2000(15)&3.816(3)&\\
\hline
Input for $f_{B_c}$: $n_f=5$\\
\hline
%\multicolumn{2}{
$\mu$[GeV]&$\alpha_s(\mu)$& $\overline{m}_b(\mu)$[GeV]&$\overline{m}_c(\mu)$[GeV]\\
3.5&0.2460(20)&4.328(8)&0.928(20)\\
Input: $\overline{m}_b(m_b)$&0.2320(20)&4.177&0.898(20)\\
5.5&0.2140(10)&3.973(2)&0.858(19)\\
6.5&0.2040(20)&3.862(2)&0.836(18)\\
7&0.2000(15)&3.816(3)&0.828(18)\\
7.5&0.1964(24)&3.775(4)&0.819(18)\\
8&0.1931(14)&3.737(4)&0.811(18)\\
9&0.1875(13)&3.672(4)&0.798(17)\\
10&0.1827(13)&3.616(5)&0.787(17)\\
11&0.1786(12)&3.567(5)&0.777(17)\\
\hline
\hline
\end{tabular}}
}
\label{tab:alfa}
\end{table}
} 
\nin
%\end{document}
%%%%%%%%%%%%%%%%%%%%%%%%%%%%%%%%%%%
%%%%%%%%%%%%%%%%%%%%%%%%%%%%%%
\section{Parametrization of the spectral function and Stability criteria}
%%%%%%%%%%%%%%%%%%%%%%%%%%%%%%
\nin

\b We shall  use the Minimal Duality Ansatz (MDA) for parametrizing the spectral function:
\bea
\frac{1}{\pi}\mbox{ Im}\psi_P(t)&\simeq& f^2_PM_P^4\delta(t-M^2_P)
  + 
  ``\mbox{QCD cont.}" \theta (t-t^P_c)~,\nnb\\
  \frac{1}{\pi}\mbox{ Im}\Pi^T_V(t)&\simeq& f^2_VM_V^2\delta(t-M^2_V)
  + 
  ``\mbox{QCD cont.}" \theta (t-t^V_c),\nnb\\
\label{eq:duality}
\eea
where $f_{P,V}$ are the decay constants defined in Eq. (\ref{eq:fp}) and the higher states contributions are smeared by the ``QCD continuum" coming from the discontinuity of the QCD diagrams and starting from a constant threshold $t^P_c,~t^V_c$ which is independent on the subtraction point $\mu$ in this standard minimal model. However, an eventual $\mu$-dependence of $t_c$ as used in some model \cite{LUCHA} should be included in the conservative range of $t_c$ used our analysis. One should notice that this MDA with constant $t_c$ describes quite well the properties of the lowest ground state as explicitly demonstrated in \cite{SNFB12a,SNFB12b} and in various examples, while it has been also successfully tested in the large $N_c$ limit of QCD in \cite{PERISa,PERISb}. 

\b Ref. \cite{SNFB12a,SNFB12b}  has explicitly tested  this simple model by confronting the predictions of the integrated spectral function within this simple parametrization with the full data measurements. One can notice  
in Figs. 1 and 2 of Ref.  \cite{SNFB12a,SNFB12b} the remarkable agreement of the model predictions and of the measured data of the $J/\psi$ charmonium and $\Upsilon$ bottomium systems for a large range of the Laplace sum rule variable $\tau$. Though it is difficult to estimate with precision the systematic error related to this simple model, this feature indicates the ability of the model for reproducing accurately the data. We expect that the same feature is reproduced for the open-charm and beauty vector meson systems where complete data are still lacking.

\b In order to extract an optimal information for the lowest resonance parameters from this rather crude  description of the spectral function and from the approximate QCD expression, one often applies the stability criteria at which an optimal result can be extracted. This stability is signaled by the existence of a stability plateau, an extremum or an inflexion point versus the changes of the external sum rule variables $\tau$ and $t_c$ where the simultaneous  requirement on the dominance over the continuum contribution and on the convergence of the OPE is satisfied. This optimization criterion demonstrated in series of papers by Bell-Bertmann \cite{BELLa,BELLb,BELLc} in the case of $\tau$ by taking the examples of harmonic oscillator and charmonium sum rules and extended to the case of $t_c$ in \cite{SNB1,SNB2} gives a more precise meaning of  the so-called ``sum rule window" originally discussed by SVZ \cite{SVZa,SVZb} and used in the sum rules literature. Similar applications of the optimization method to the pseudoscalar $D$ and $B$ open meson states have been successful when compared with results from some other determinations as discussed in Ref.\,\cite{SNFB12a,SNFB12b} and reviewed in\,\cite{SNB1,SNB2,SNREV14} and in some other recent reviews\,\cite{ROSNERa,ROSNERb,LATT13} quoted in the present paper. 

\b In this paper, we shall add to the previous well-known stability criteria, the one associated  to the requirement of stability  versus the arbitrary subtraction constant $\mu$ often put by hand  in the current literature  and which is often the source of large errors from the PT series in the sum rule analysis.  Indeed, the choice of the region of variation of $\mu$ is not always well founded like e.g taking $\mu$ between 1.3 and 3 GeV \cite{PIVOV} and between 1 to 3 GeV \cite{LUCHA} in the case of the $D^*$ meson or by taking a ``default" value of $\mu\simeq 3$ GeV \cite{PIVOV} for evaluating the central value of $f_{B^*}$. The $\mu$-stability procedure has been applied recently in\,\cite{SNFB12a,SNFB12b,SNFB13,SNLIGHT,SNREV14}\,\footnote{Some other alternative approaches for optimizing the PT series can be found in \cite{STEVENSONa,STEVENSONb,STEVENSONc,STEVENSONd,STEVENSONe}.} which gives a much better meaning on the choice of $\mu$-value at which the observable is extracted, while the errors  in the determinations of the results have been reduced due to a better control of the $\mu$ region of variation which is not the case in the existing literature.
% previous sum rule results based on the arbitrary choice of $\mu$. 
%%%%%%%%%%%%%%%%%%%%%%%%%%%%%%
\section{The decay constant  $f_{D^*}$}
%%%%%%%%%%%%%%%%%%%%%%%%%%%%%%
\nin
\subsection{ The ratio $f_{D^*}/f_{D}$}
\begin{figure}[hbt] 
\begin{center}
\centerline {\hspace*{-8.5cm} a) }\vspace{-0.6cm}
{\includegraphics[width=8.5cm]{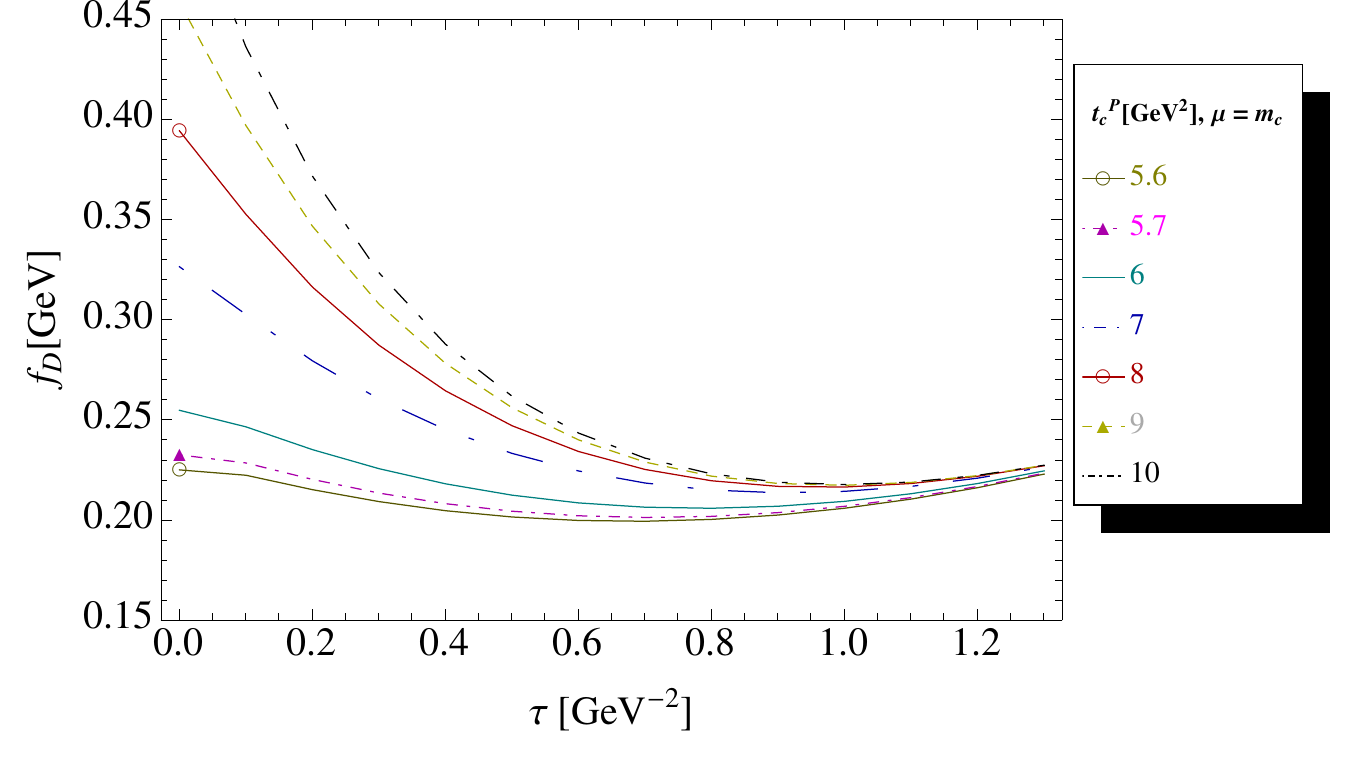}}
\centerline {\hspace*{-8.5cm} b) }\vspace{-0.3cm}
{\includegraphics[width=8.5cm]{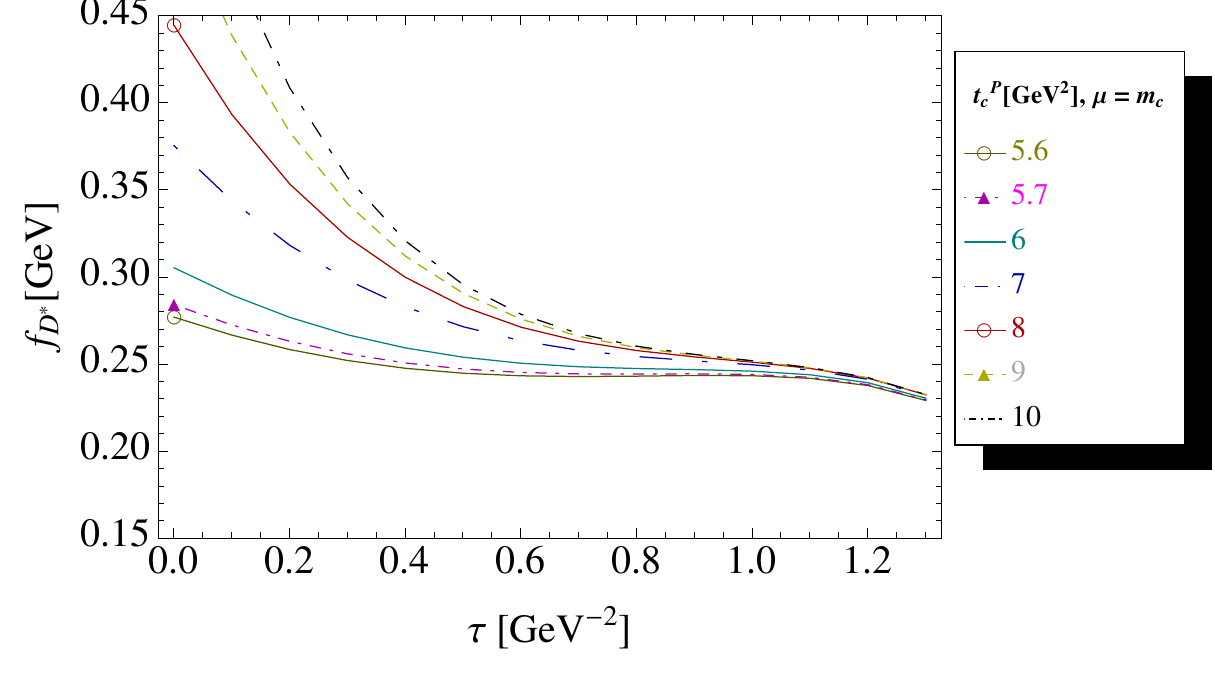}}
\caption{
\scriptsize 
{\bf a)} $\tau$-behaviour of $f_{D}$  from  ${\cal L}_{P} $ for different values of $t_c$, at a given value of the subtraction point $\mu=m_c$; {\bf b)} the same as in a) but for $f_{D^*}$ from ${\cal L}_{V} $.}
\label{fig:fdtau} 
\end{center}
\end{figure}

\begin{figure}[hbt] 
\begin{center}
{\includegraphics[width=8cm  ]{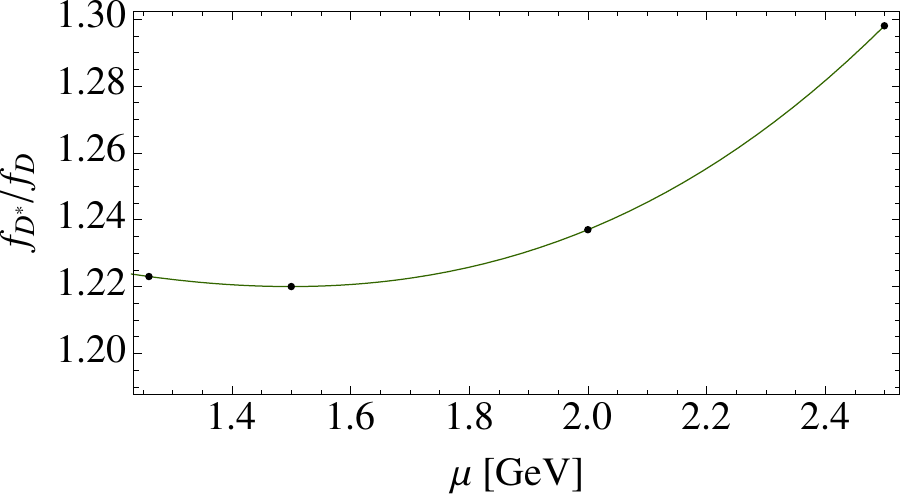}}
\caption{
\scriptsize 
Values of  $f_{D^*}/f_D$  at different values of the subtraction point $\mu$. }
% The dashed horizontal lines are the values if one takes the errors from the best determination.}
\label{fig:fd*fdmu} 
\end{center}
\end{figure}

\nin
%%%%%%%%%%%%%%%%%%%%%%%%%%%
We start by showing in Fig. \ref{fig:fdtau}, the $\tau$-behaviour of the decay constants $f_{D^*}$ and $f_D$ at given value of the subtraction point $\mu=m_c$ for different values of the continuum threshold $t_c$. We have assumed that :
\beq
\sqrt{t_c^{D^*}}-\sqrt{t_c^D}\simeq M_{D^*}-M_{D}=140.6~{\rm MeV}.
\eeq
 for the vector and pseudoscalar channels.
% \,\footnote{Neglecting this difference between $t^V_c$ and $t_c^P$ would induce a negligible error of 0.6\% in the ratio of decay constants.}. 
 For the pseudoscalar channel,  we have used the expression in Eq. (20) of \cite{SNFB12a,SNFB12b} consistently truncated at the same order of PT and NP series as  the one in Eq. (\ref{eq:vec}) for the vector channel. One can notice in Fig. \ref{fig:fdtau} that working directly with the ratio in Eq. (\ref{eq:ratio})
%\beq
%{\cal R}_{V/P}\equiv {{\cal L}_V\over {\cal L}_P}~,
%\eeq
by taking the same value $\tau_V=\tau_P$ is inaccurate as the two sum rules ${\cal L}_V(\tau)$ and ${\cal L}_P(\tau)$ are not optimized at the same value of $\tau$ (minimum for $f_D$ and inflexion point for $f_{D^*}$). Therefore, for a given value of $t_c$, we take separately the value of each sum rule at the corresponding value of $\tau$
where they present minimum and/or inflexion point and then take their ratio. For a given $\mu$, the optimal result corresponds to the mean obtained in range of values of $t_c$ where one starts to have a $\tau$-stability  ($t_c\simeq 5.6-5.7$ GeV$^2$ for $\tau\simeq 0.6$ GeV$^{-2}$) and a $t_c$-stability ($t_c\simeq 9.5\sim 10.5$ GeV$^2$ for $\tau\simeq 0.8$ GeV$^{-2}$). 
Now, we look for the $\mu$-stability by plotting versus $\mu$ the previous optimal ratio $f_{D^*}/f_D$ in the variables $\tau$ and $t_c$. The results are shown in Fig. \ref{fig:fd*fdmu}. We obtain a minimum for $\mu=(1.5\pm 0.1)$ GeV which is about the average 1.5 GeV of \cite{PIVOV} and 1.84 GeV used in \cite{LUCHA}. At this minimum, we deduce the final result:
\bea
f_{D^*}/f_D&=&1.218(6)_{t_c}(27)_{\tau}(23)_{svz}(4)_\mu\nnb\\
&=&1.218(36)
\label{eq:fdd*}
\eea
where the error from the QCD expression within the SVZ expansion is the quadratic sum of
\bea
(23)_{svz}&=&(7)_{\alpha_s}(2)_{\alpha_s^3}(3)_{m_c}(0)_{\la\bar dd\ra}(18)_{\la \alpha_sG^2\ra}\nnb\\
&&(12)_{\la\bar dGd\ra}(0)_{\la g^3G^3\ra}(1)_{\la\bar dd\ra^2}~.
\eea
To be more conservative, we have multiplied by a factor 2 the error due to the choice of the subtraction point $\mu$. 
One can notice that the largest error comes from $\tau$ which is due to the inaccurate localization of the inflexion point. The error due to $t_c$ is smaller as expected in the determination of the ratio which is not the case for the direct extraction of the decay constants. The errors due to ${\la \alpha_sG^2\ra}$ and ${\la\bar dGd\ra}$ are large due to the opposite sign of their contributions in the vector and pseudoscalar channels which add when taking the ratio. 
%The final results for $f_{D^*}/f_D$ with their corresponding errors  are given in Table \ref{tab:fdmu} for different values of $\mu$. One can notice in this table that the dominant error in the determination of the ratio comes from our conservative range of values of the continuum threshold $t_c$ while the ones from the QCD expression are negligible. One may choose (like done in the current literature) its value at the physical threshold but this choice maybe misleading as the QCD continuum smears all higher radial excited state contributions. 
%%%%%%%%%%%%%%%%%%%%%%%%%
%\vspace*{-0.3cm}
\nin
%%%%%%%%%%%%%%%%%%%%%%%%
%%%%%%%%%%%%%%%%%%%%%%%%%%%%%%
\subsection{Estimate of $f_{D^*}$}
\begin{figure}[hbt] 
\begin{center}
{\includegraphics[width=8cm  ]{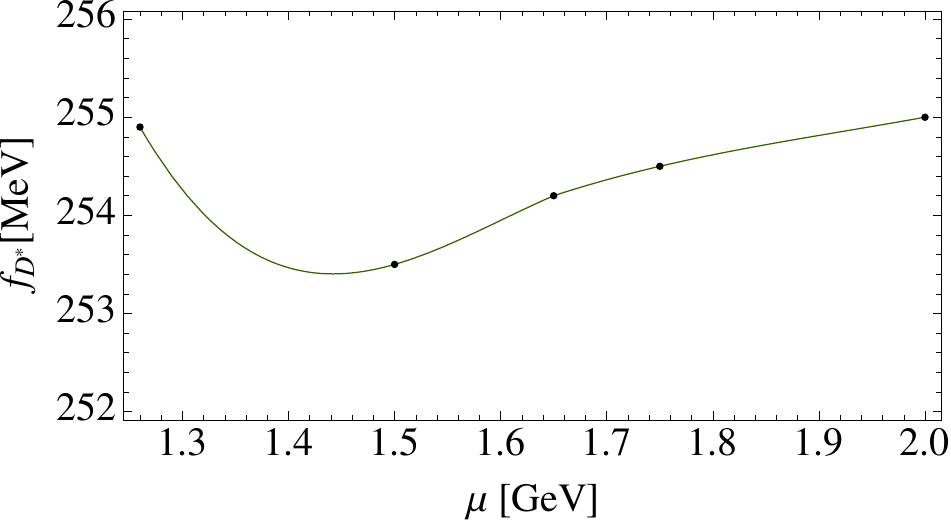}}
\caption{
\scriptsize 
Values of  $f_{D^*}$  at different values of the subtraction point $\mu$. }
% The dashed horizontal lines are the values if one takes the errors from the best determination.}
\label{fig:fd*mu} 
\end{center}
\end{figure}

\nin
%%%%%%%%%%%%%%%%%%%%%%%%%%%
Using the value $f_D=204(6)$ MeV (Table 8 of \cite{SNFB12a,SNFB12b}) obtained under a similar strategy and the ratio in Eq. (\ref{eq:fdd*}), we deduce:
\beq
f_{D^*}=248.5(10.4)~{\rm MeV}~,
\label{eq:fd*ratio}
\eeq
where the errors have been added quatratically. \\
One can also extract directly $f_{D^*}$ using the analysis of Fig. \ref{fig:fdtau}b and a similar strategy as the one used for extracting the ratio $f_{D^*}/f_D$. The $\mu$ behaviour of the optimal $\tau$ and $t_c$ result
is given in Fig. \ref{fig:fd*mu}. A minimum is also obtained for $\mu\simeq 1.5$ GeV at which we deduce:
\bea
f_{D^*}&=&253.5(11.5)_{t_c}(5.7)_{\tau}(13)_{svz}(1)_\mu~{\rm MeV}\nnb\\
&=&253.5(18.3)~{\rm MeV},
\label{eq:fd*1}
\eea
with:
\bea
(13)_{svz}&=&(0.5)_{\alpha_s}(12.3)_{\alpha_s^3}(0.6)_{m_c}(3.8)_{\la\bar dd\ra}(1.8)_{\la \alpha_sG^2\ra}\nnb\\
&&(1.4)_{\la\bar dGd\ra}(0.4)_{\la g^3G^3\ra}(0.4)_{\la\bar dd\ra^2}~,
\eea
where again the error due to $\mu$ has been mutiplied by a factor 2 for a more conservative error\,\footnote{In the remaining part of the paper, we shall systematically multiply the error due to $\mu$ by a factor 2 for a conservative estimate of this source of error. The original errors due to $t_c$ and $\tau$ are already conservative because the associated stability criteria  correspond to large ranges of these parameters.}.
Our final result will be the mean of the two determinations in Eqs. (\ref{eq:fd*ratio}) and (\ref{eq:fd*1}) which is:
\bea
\la f_{D^*}\ra&=&249.7(10.5)(1.2)_{syst}~{\rm MeV}\nnb\\
&=&250(11)~{\rm MeV}~,
\lb{eq:fd*}
\eea
where the 1st error comes from the most precise determination and the 2nd one from the distance of the mean value to it.
This result  is inside the range of the recent sum rules results from \cite{PIVOV,LUCHA} but  lower than the unique available lattice value \cite{BECIR}: $f_{D^*}=278(16)$ MeV where an independent estimate from some other lattice groups is required.  
  %%%%%%%%%%%%%%%%%%%%%%%%%
\nin
  %%%%%%%%%%%%%%%%%%%%%%%%%
\nin

%%%%%%%%%%%%%%%%%%%%%%%%%%%%%%%
%%%%%%%%%%%%%%%%%%%%%%%%%%%
\subsection{Comparison of the errors with other estimates}
\nin
%%%%%%%%%%%%%%%%%%%%%%%%%%%
It is informative to compare the errors obtained here with the ones from \cite{PIVOV} which use the same MDA model of the spectral than in this paper.  

\b One can obtain from Table~2 of this paper:
\bea
f_{D^*}=241.9(6.2)_{t_c}(^{+3.6}_{-5.0})_{\tau}(9.8)_{svz}(^{+17.3}_{-3.9})_\mu~{\rm MeV}~,
\label{eq:fd*comp}
\eea
compared with our previous estimate in Eq. \ref{eq:fd*1}.  One can notice that, for a given subtraction point $\mu=1.5$ GeV, the errors from our sum rule analysis are systematically slightly larger than the ones in Ref. \cite{PIVOV} indicating that our optimizations of the $\tau$-sum rule and $t_c$-continuum variables reproduce almost the same results as the one using a slightly different approach. The reason is that we have considered a more conservative range of $\tau$ and $t_c$ values than the ones in \cite{PIVOV}.  The good agreement of the results (indirectly) supports our optimization procedures. 

\b The main source of errors in \cite{PIVOV} is the arbitrary choice of the $\mu$ subtraction region which is strongly constrained in your optimization of the value of $\mu$. In fact, if we move $\mu$ from 1.5 to 2.5 GeV, $f_{D^*}$ increases by about 12 MeV which is comparable with the value 17 MeV obtained in Ref. \cite{PIVOV} by taking $\mu$ from 1.5 to 3 GeV. The small variation of  $f_{D^*}$ around the $\mu$ stability point explains the relatively small error in Eq. (\ref{eq:fd*1}) compared to the ones in Ref. \cite{PIVOV} and in Ref. \cite{LUCHA} (which uses a slightly different model for the QCD continuum), where $\mu$ has been varied in a larger range from 1 to 3 GeV. This choice is not justified by our optimization procedure where only  values of $\mu$ around the minimum in Fig. \ref{fig:fd*mu} should be considered. We shall see later on that a similar feature occurs for the estimate of the other decay constants. 
%%%%%%%%%%%%%%%%%%%%%%%%%%%
\subsection{Upper bound on $f_{D^*}$}
\begin{figure}[hbt] 
\begin{center}
{\includegraphics[width=10cm  ]{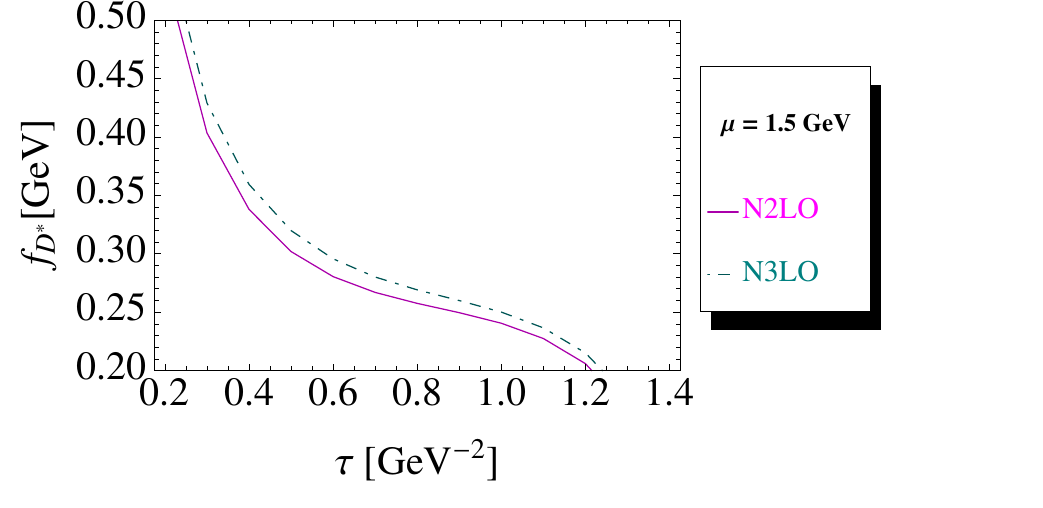}}\\
\caption{
\scriptsize 
Upper bounds on $f_{D^*}$ (upper curves)  at different values of $\tau$ for a given value $\mu=$1.5 GeV of the subtraction point $\mu$. One can notice a good convergence of the PT series by comparing the calculated N2LO and estimated N3LO terms.} 
\label{fig:fd*bound_tau} 
\end{center}
\end{figure}

\begin{figure}[hbt] 
\begin{center}
{\includegraphics[width=8cm  ]{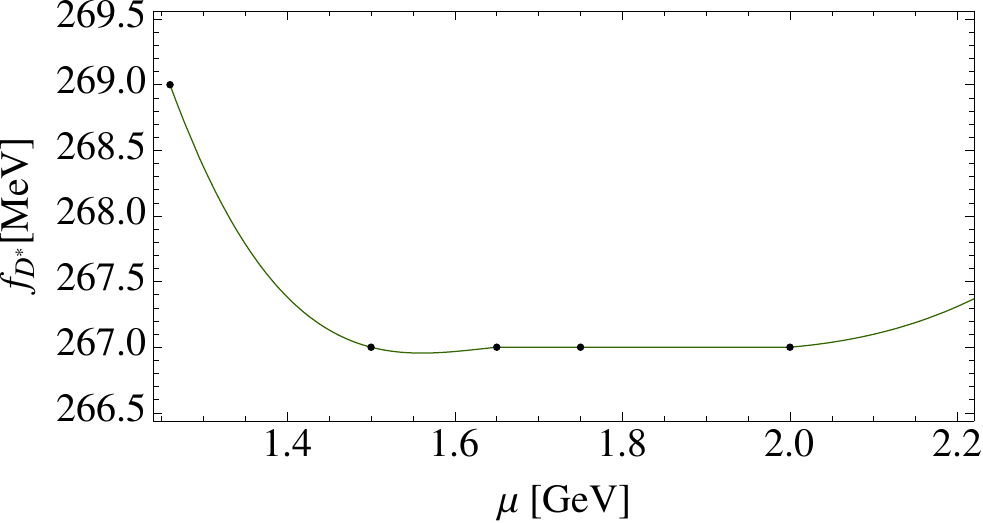}}
\caption{
\scriptsize 
Upper bound of $f_{D^*}$  versus the values of the subtraction point $\mu$. }
\label{fig:fd*bound_mu} 
\end{center}
\end{figure}

\nin
%%%%%%%%%%%%%%%%%%%%%%%%%%%
We derive an upper bound on $f_{D^*}$ by considering the positivity of the QCD continuum contribution to the spectral function
and by taking the limit where $t_c\to\infty$ in Eq. (\ref{eq:lsr}) which corresponds to a full saturation of the spectral function by the lowest ground state contribution. The result of the analysis versus the change of $\tau$ for a given value of $\mu=1.5$ GeV is given in Fig.~\ref{fig:fd*bound_tau} where one can observe like in the previous analysis the presence of a $\tau$-inflexion point. We also show in this figure the good convergence of the PT series
by comparing the result at N2LO and the one including an estimate of the N3LO term based on the geometric growth of the PT coefficients. We show in Fig \ref{fig:fd*bound_mu} the variation of the optimal bound versus the subtraction point $\mu$ where we find a  region of $\mu$ stability from 1.5 to  2 GeV. 
  %%%%%%%%%%%%%%%%%%%%%%%%%
\nin
%%%%%%%%%%%%%%%%%%%%%%%%%%%%%%%
We obtain:
\beq
f_{D^*}\leq 267(10)_\tau(14)_{svz}(0)\mu~{\rm MeV}~,
\label{eq:fdbound1}
\eeq
with:
\bea
(14)_{svz}&=&(1.4)_{\alpha_s}(13)_{\alpha_s^3}(1.3)_{m_c}(3)_{\la\bar dd\ra}(3)_{\la \alpha_sG^2\ra}\nnb\\
&&(2.5)_{\la\bar dGd\ra}(0)_{\la g^3G^3\ra}(0.7)_{\la\bar dd\ra^2}~.
\eea

 Alternatively, we combine the upper bound $f_D\leq 218.4(1.4)$ MeV obtained \cite{SNFB12a,SNFB12b} with the previous ratio in Eq. (\ref{eq:fdd*}) and deduce:
\beq
f_{D^*}\leq 266(8)~{\rm MeV}~,
\label{eq:fdbound2}
\eeq
where we have added the errors quadratically. The good agreement of the results in Eqs. (\ref{eq:fdbound1}) and (\ref{eq:fdbound2}) indicates
the self-consistency of the approaches. This bound is relatively strong compared to the estimate in Eq. (\ref{eq:fd*}) while  
the recent lattice estimate  $f_{D^*}\simeq 278(16)~{\rm MeV}$ obtained in \cite{BECIR} is at its borderline. 
  %%%%%%%%%%%%%%%%%%%%%%%%%
\nin
%%%%%%%%%%%%%%%%%%%%%%%%%%%%%%%
%%%%%%%%%%%%%%%%%%%%%%%%%%%%%%
\section{The decay constant  $f_{B^*}$ }
%%%%%%%%%%%%%%%%%%%%%%%%%%%%%%
\nin
%%%%%%%%%%%%%%%%%%%%%%%%%
%\vspace*{-0.3cm}
\begin{figure}[hbt] 
\begin{center}
\centerline {\hspace*{-8.5cm} a) }\vspace{-0.6cm}
{\includegraphics[width=9cm  ]{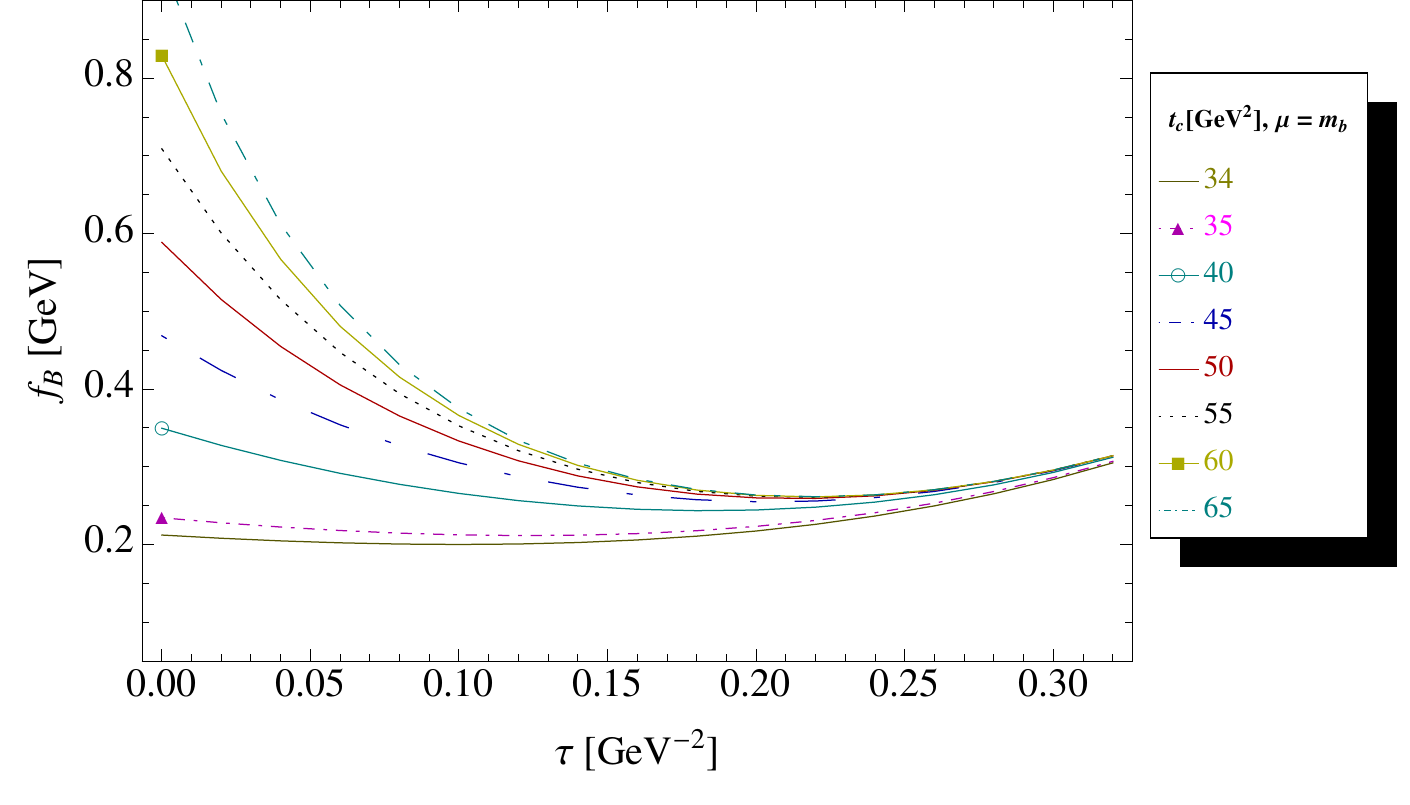}}
\centerline {\hspace*{-8.5cm} b) }\vspace{-0.3cm}
{\includegraphics[width=9cm  ]{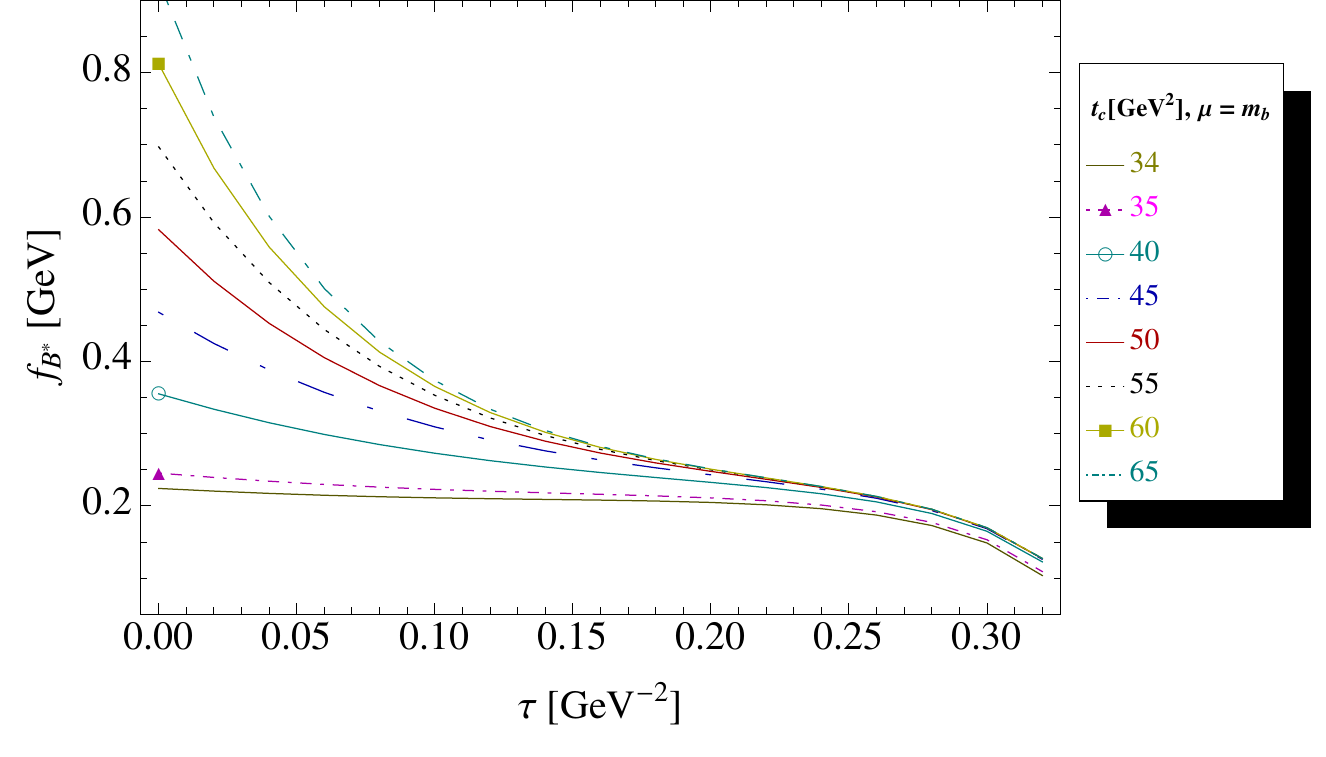}}
\caption{
\scriptsize 
{\bf a)} $\tau$-behaviour of $f_{B}$  from  ${\cal L}_{P} $ for different values of $t_c$, at a given value of the subtraction point $\mu=m_b$; {\bf b)} the same as in a) but for $f_{B^*}$ from ${\cal L}_{V} $.}
\label{fig:fbtau} 
\end{center}
\end{figure} 
\nin
%%%%%%%%%%%%%%%%%%%%%%%%%
We extend the analysis to the case of the $B^*$ meson. We use the set of parameters in Table \ref{tab:param} and  \ref{tab:alfa}. We show the $\tau$-behaviour of $f_{B}$ and $f_B^*$ in Fig. \ref{fig:fbtau}, where the shape is similar to the case of $f_{D^*}$ and $f_D$. 

%%%%%%%%%%%%%%%%%%%%%%%%%%%%%%
\subsection{Estimate of  $f_{B^*}$}
\begin{figure}[hbt] 
\begin{center}
{\includegraphics[width=8cm  ]{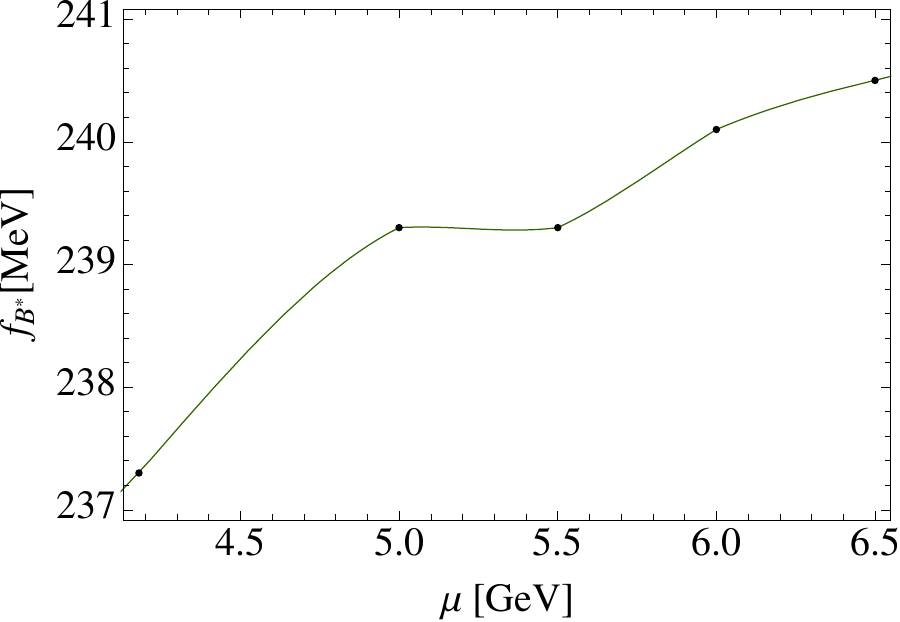}}
\caption{
\scriptsize 
Values of  $f_{B^*}$  at different values of the subtraction point $\mu$. }
\label{fig:fb*mu} 
\end{center}
\end{figure}

\nin
%%%%%%%%%%%%%%%%%%%%%%%%%%%
Using the previous strategy,  we estimate  $f_{B^*}$ from the analysis in Fig. \ref{fig:fbtau}b where the $\tau$-stability is reached from $t_c=34$ GeV$^2$ while the $t_c$ stability starts from $t_c=(55-60)$ GeV$^2$. We show the $\mu$ behaviour of the optimal
result in Fig. \ref{fig:fb*mu} where we find a clear inflexion point for $\mu=(5.0-5.5)$ GeV at which we extract the optimal result:
\bea
f_{B^*}&=&239(38)_{t_c}(1)_{\tau}(2.7)_{svz}(1.4)_\mu~{\rm MeV}\nnb\\
&=&239(38)~{\rm MeV},
\label{eq:fb*mu}
\eea
with:
\bea
(2.7)_{svz}&=&(0.7)_{\alpha_s}(2)_{\alpha_s^3}(0.4)_{m_b}(1.6)_{\la\bar dd\ra}(0.4)_{\la \alpha_sG^2\ra}\nnb\\
&&(0.3)_{\la\bar dGd\ra}(0)_{\la g^3G^3\ra}(0)_{\la\bar dd\ra^2}~,
\eea
where the error in $\mu$ comes by taking $\mu=(5.5\pm 0.5)$ GeV.
  %%%%%%%%%%%%%%%%%%%%%%%%%
\nin
  %%%%%%%%%%%%%%%%%%%%%%%%%
%%%%%%%%%%%%%%%%%%%%%%%%
\subsection{The ratio $f_{B^*}/f_{B}$}
\begin{figure}[hbt] 
\begin{center}
{\includegraphics[width=8cm  ]{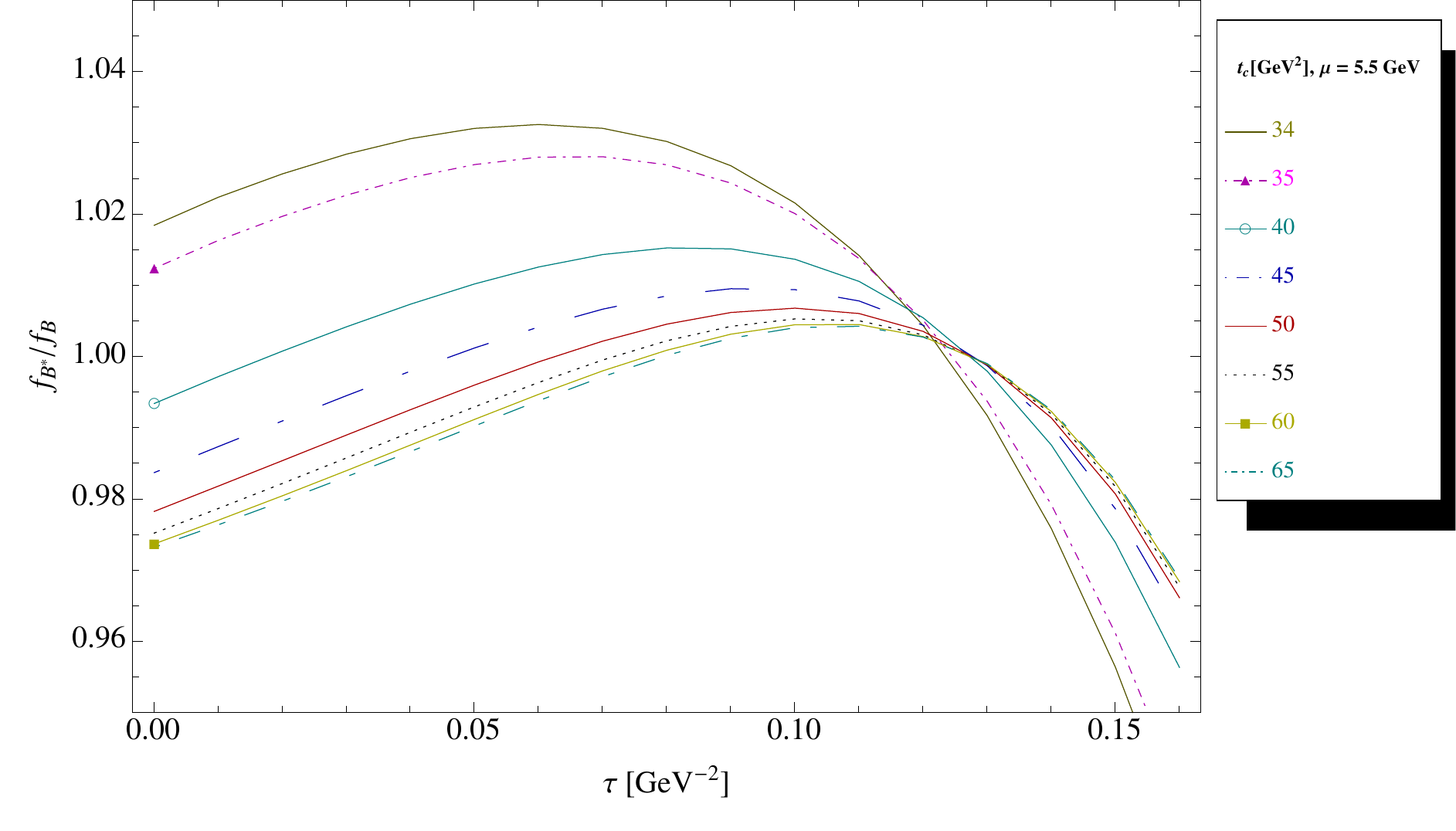}}
\caption{
\scriptsize 
$\tau$-behaviour of $f_{B^*}/f_B$ for different values of $t_c$, at a given value of the subtraction point $\mu=5.5$ GeV}
\label{fig:rbst_tau} 
\end{center}
\end{figure}

\begin{figure}[hbt] 
\begin{center}
{\includegraphics[width=8cm  ]{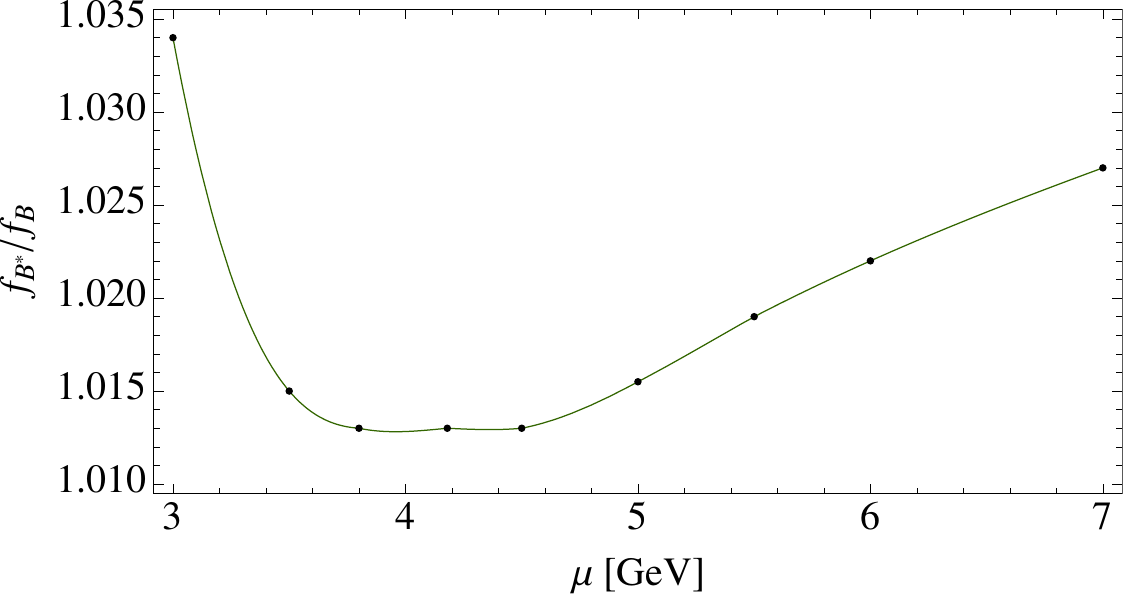}}
\caption{
\scriptsize 
Values of  $f_{B^*}/f_B$  at different values of the subtraction point $\mu$. }
\label{fig:rbst_mu} 
\end{center}
\end{figure}

\nin
%%%%%%%%%%%%%%%%%%%%%%%%%%%
Here, we extract  directly the ratio $f_{B^*}/f_B$ from the ratio of sum rules. We show in Fig. \ref{fig:rbst_tau} its $\tau$ 
behaviour for different values of $t_c$ for $\mu=5.5$ GeV from which we deduce as optimal value the mean of the $\tau$-minima obtained  from $t_c=34$ to 60 GeV$^2$. We show  in Fig. \ref{fig:rbst_mu} the $\mu$ behaviour of these optimal results where we find a minimum in $\mu$ around 3.8-4.5 GeV and a slight inflexion point around 5.5 GeV. We consider as a final result the mean of the ones from these two regions of $\mu$:
\bea
f_{B^*}/f_{B}&=&1.016(12)_{t_c}(1)_\tau(9)_{svz}(6)_\mu\nnb\\
&=&1.016(16)~,
\lb{eq:rb*mu}
\eea
with:
\bea
(9)_{svz}&=&(3)_{\alpha_s}(6)_{\alpha_s^3}(3)_{m_b}(4)_{\la\bar dd\ra}(3)_{\la \alpha_sG^2\ra}\nnb\\
&&(1)_{\la\bar dGd\ra}(1)_{\la g^3G^3\ra}(1)_{\la\bar dd\ra^2}~.
\eea
Our results in Eqs. (\ref{eq:fb*mu}) and (\ref{eq:rb*mu}) are comparable with the ones in \cite{PIVOV} but with large errors due mainly to our conservative range of $t_c$ values. However, the value of $\mu$ at which our optimal results are obtained does not favour the choice $\mu=3$ GeV adopted in \cite{PIVOV}. 
Combining the results in Eq. (\ref{eq:rb*mu}) with the value $f_B=206(7)$ MeV obtained in \cite{SNFB12a,SNFB12b}, we deduce:
\beq
f_{B^*}=209(8)~{\rm MeV}~,
\lb{eq:fb*}
\eeq
which is more accurate than the direct determination in Eq. (\ref{eq:fb*mu}) and where the main error comes from the one of $f_B$ extracted in \cite{SNFB12a,SNFB12b}. We consider the result in Eq. (\ref{eq:fb*}) which is also the mean of the results in Eqs. (\ref{eq:fb*}) and (\ref{eq:fb*mu}) as our final determination. 
%%%%%%%%%%%%%%%%%%%%%%%%%
%\vspace*{-0.3cm}
\nin
%%%%%%%%%%%%%%%%%%%%%%%%%
  %%%%%%%%%%%%%%%%%%%%%%%%%
\nin
  %%%%%%%%%%%%%%%%%%%%%%%%%
  %%%%%%%%%%%%%%%%%%%%%%%%%
\begin{figure}[hbt] 
\begin{center}
{\includegraphics[width=9.5cm  ]{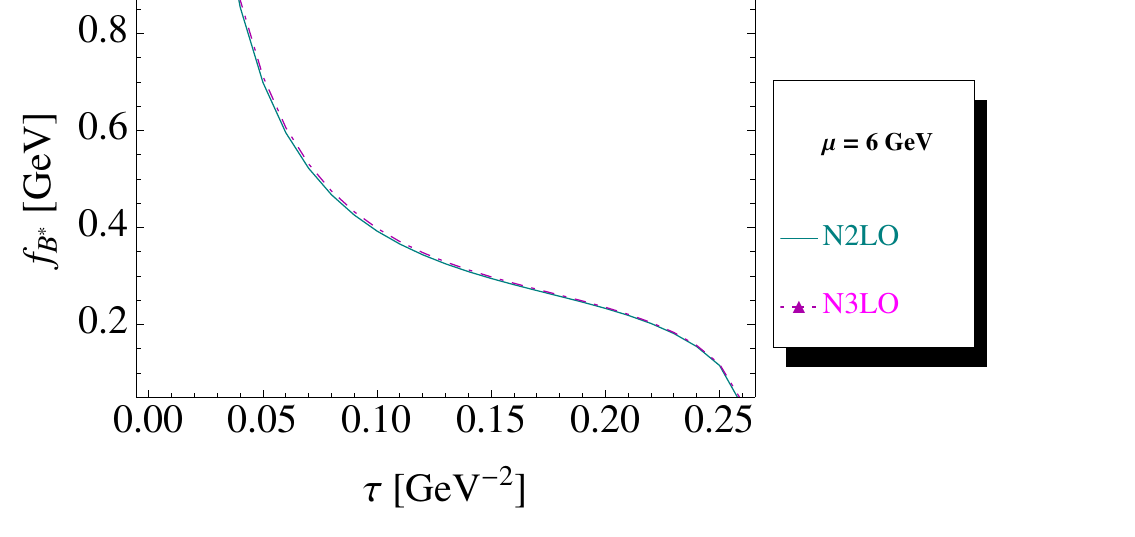}}
\caption{
\scriptsize 
$\tau$-behaviour of the upper bound of $f_{B^*}$  from  ${\cal L}_{P} $ at a given value of the subtraction point $\mu=6$ GeV for N2LO and N3LO truncation of the PT series.}
\label{fig:fbbound_tau} 
\end{center}
\end{figure} 
\nin
%%%%%%%%%%%%%%%%%%%%%%%%%%%%%%%
  %%%%%%%%%%%%%%%%%%%%%%%%%
\begin{figure}[hbt] 
\begin{center}
{\includegraphics[width=8.5cm  ]{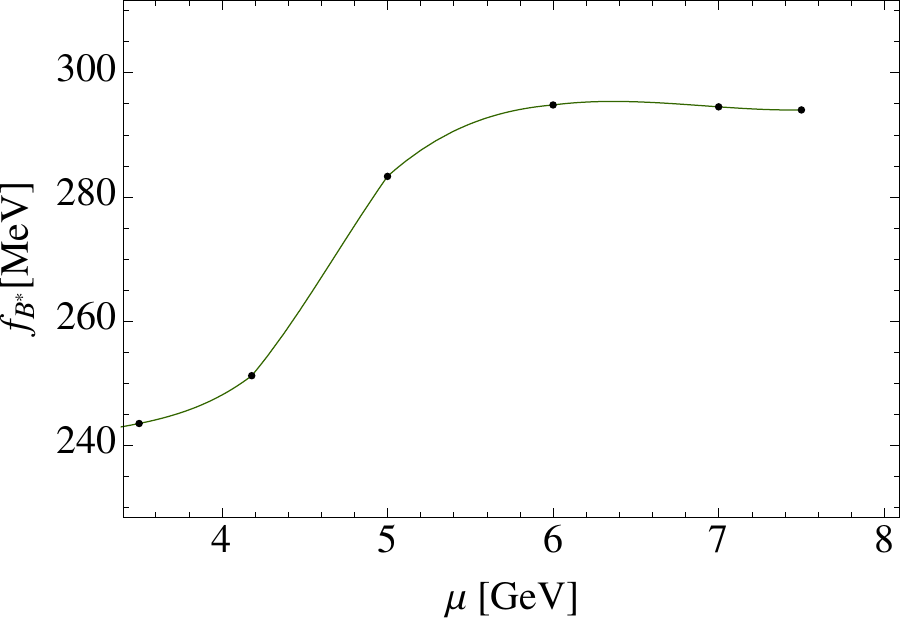}}
\caption{
\scriptsize 
 $\mu$-behaviour of the upper bound of $f_{B^*}$.}
\label{fig:fbbound_mu} 
\end{center}
\end{figure} 
\nin
%%%%%%%%%%%%%%%%%%%%%%%%%%%%%%%
%%%%%%%%%%%%%%%%%%%%%%%%%%%
\subsection{Upper bound on $f_{B^*}$}
\nin
%%%%%%%%%%%%%%%%%%%%%%%%%%%
Like in the case of the $D^*$ meson, we extract directly an upper bound on $f_{B^*}$ by using the positivity of the QCD continuum to the spectral function. We show the $\tau$-behaviour of this upper bound in Fig. \ref{fig:fbbound_tau} and the $\mu$ behaviour of the optimal bound in Fig. \ref{fig:fbbound_mu}.
We deduce the optimal bound at $\mu=(6\pm 0.5)$ GeV:
\bea
f_{B^*}&\leq& 295(14)_\tau (4)_{svz}(10)_\mu~{\rm MeV}\nnb\\
&\leq&295(18)~{\rm MeV}~.
\label{eq:fb*_bound}
\eea
%which is rather weak unlike the case of the $D^*$ meson.\\
We consider the previous values of $f_{D^*}$ and $f_{B^*}$ as improvement of our earlier results in \cite{SNB2} and \cite{SNFB88}. The results indicate that in the heavy quark limit one has $f_B\simeq f_{B^*}$ which indicates a good realization of the heavy quark symmetry as expected from HQET \cite{NEUBERT2}, while $m_c$ is still too low at which a such symmetry is broken by the charm quark mass and QCD radiative corrections. 
%%%%%%%%%%%%%%%%%%%%%%%%%%%%%%%%%%%%%%
\section{ SU(3) breaking for $f_{D^*_s}$ and $f_{D^*_s}/f_{D^*}$ }
\begin{figure}[hbt] 
\begin{center}
{\includegraphics[width=9cm  ]{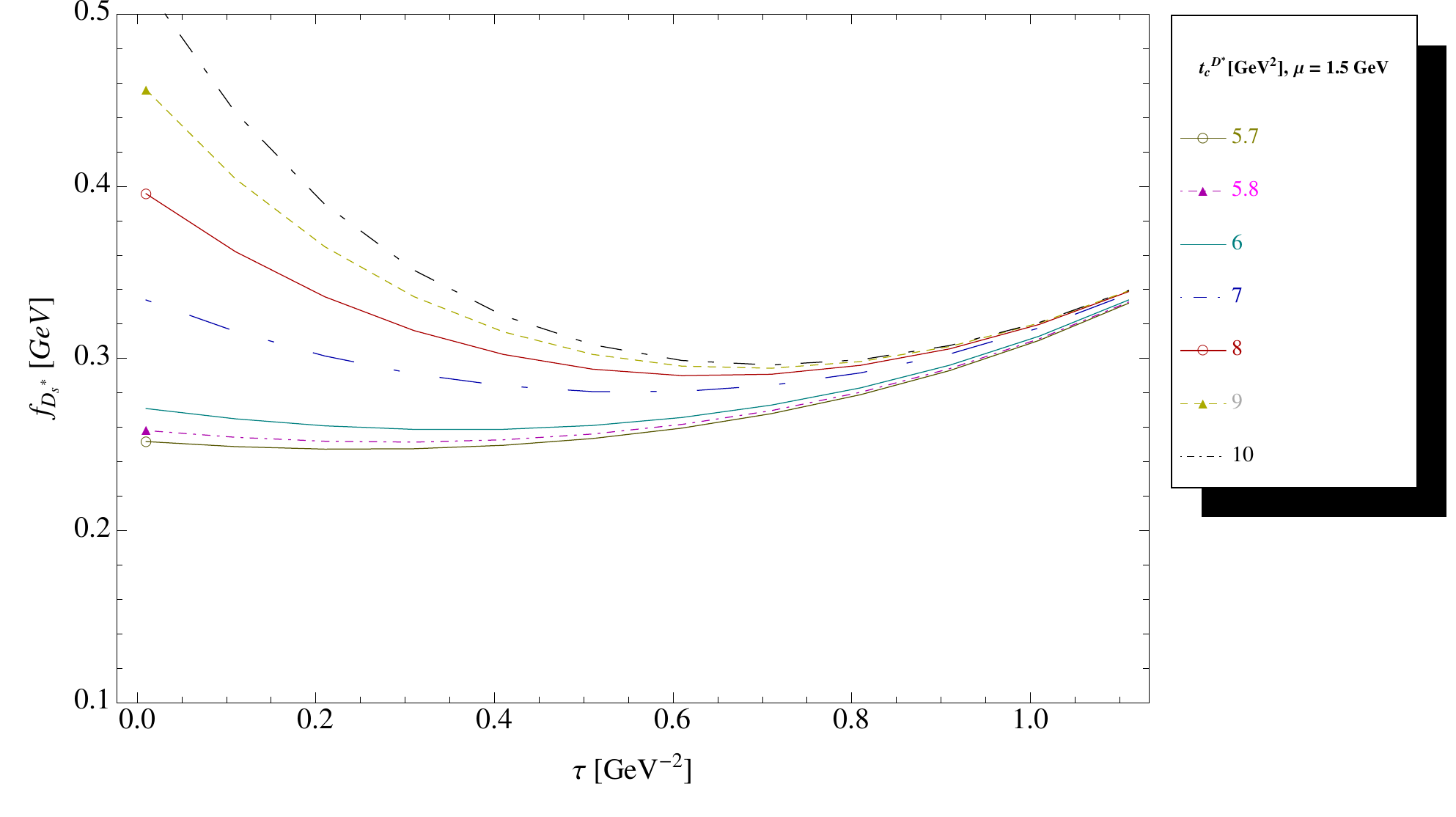}}
\caption{
\scriptsize 
 $\tau$-behaviour of $f_{D^*_s}$  from  ${\cal L}_{V} $ for different values of $t_c$, at a given value of the subtraction point $\mu=$1.5 GeV.}
\label{fig:fd*stau} 
\end{center}
\end{figure}

\begin{figure}[hbt] 
\begin{center}
{\includegraphics[width=8cm  ]{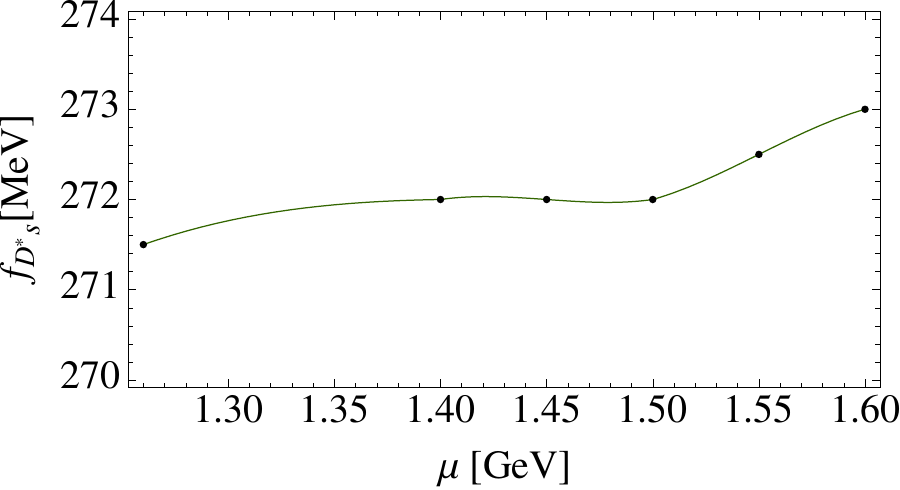}}
\caption{
\scriptsize 
 $\mu$-behaviour of $f_{D^*_s}$.}
\label{fig:fd*smu} 
\end{center}
\end{figure}

%%%%%%%%%%%%%%%%%%%%%%%%%%%%%%%%%%%%%%
We pursue the same analysis for studying the $SU(3)$ breaking for  $f_{D^*_s}$ and the ratio $f_{D^*_s}/f_{D^*}$. We work with the complete massive $(m_s\not=0)$ LO expression of the PT spectral function obtained in \cite{FNR} and the massless $(m_s=0)$ expression known to N2LO used in the previous sections.  We include the NLO PT corrections due to linear terms in $m_s$ obtained in \cite{PIVOV}.   We show the $\tau$ behaviour of different results in Fig.~\ref{fig:fd*stau} for a given value of $\mu=1.5$ GeV and for different $t_c$. We study the $\mu$ dependence of these results in Fig.~\ref{fig:fd*smu} where a nice $\mu$ stability is reached for $\mu\simeq 1.4-1.5$ GeV. 
%%%%%%%%%%%%%%%%%%%%%%%%%
%\vspace*{-0.3cm}
\nin
%\vspace*{-0.3cm}
%%%%%%%%%%%%%%%%%%%%%%%%%
\nin
%%%%%%%%%%%%%%%%%%%%%%%%%
We have used :
\beq
\sqrt{t_c^{D^*_s}}-\sqrt{t_c^{D^*}}=M_{D^*_s}-M_{D^*}=102~{\rm MeV}~.
\eeq
Taking the conservative result ranging from the beginning of $\tau$-stability ($t_c\simeq 5.7$ GeV$^2$) until the beginning of $t_c$-stability of about (9--10) GeV$^2$, we obtain at $\mu$=1.5 GeV:
\bea
f_{D^*_s}&=&272(24)_{t_c}(2)_\tau(18)_{svz}(2)_\mu~{\rm MeV}\nnb\\
&=& 272(30)~{\rm MeV}~,
\lb{eq:fd*s}
\eea
with:
\bea
(18)_{svz}&=&(0)_{\alpha_s}(14)_{\alpha_s^3}(1)_{m_c}(2)_{\la\bar dd\ra}(1.5)_{\la \alpha_sG^2\ra}\nnb\\
&&(0.8)_{\la\bar dGd\ra}(0)_{\la g^3G^3\ra}(0)_{\la\bar dd\ra^2}\nnb\\
&&(0.3)_{m_s}(2)\kappa~.
\eea
Taking the PT linear term in $m_s$ at lowest order and $t_c=7.4$ GeV$^2$, we obtain $f_{D^*_s}=291$ MeV in agreement with the one 293 MeV of \cite{PIVOV} obtained in this way. The inclusion of the complete LO term 
decreases this result by about 5 MeV while the inclusion of the NLO PT $SU(3)$ breaking terms increases the result by about the same amount. However, we do not see any justification for choosing the value of $t_c=7.4$ GeV$^2$ used in \cite{PIVOV}. Instead, one can consider that our result coming from the mean of the one at $t_c=5.7$ and 10 GeV$^2$ is more conservative. Combining the result in Eq. (\ref{eq:fd*s}) with the one in Eq. (\ref{eq:fd*}), we deduce the ratio:
\beq
f_{D^*_s}/ f_{D^*}=1.090(70)~,
\lb{eq:rd*stau1}
\eeq
%%%%%%%%%%%%%%%%%%%%%%%%%
\begin{figure}[hbt] 
\begin{center}
{\includegraphics[width=8cm  ]{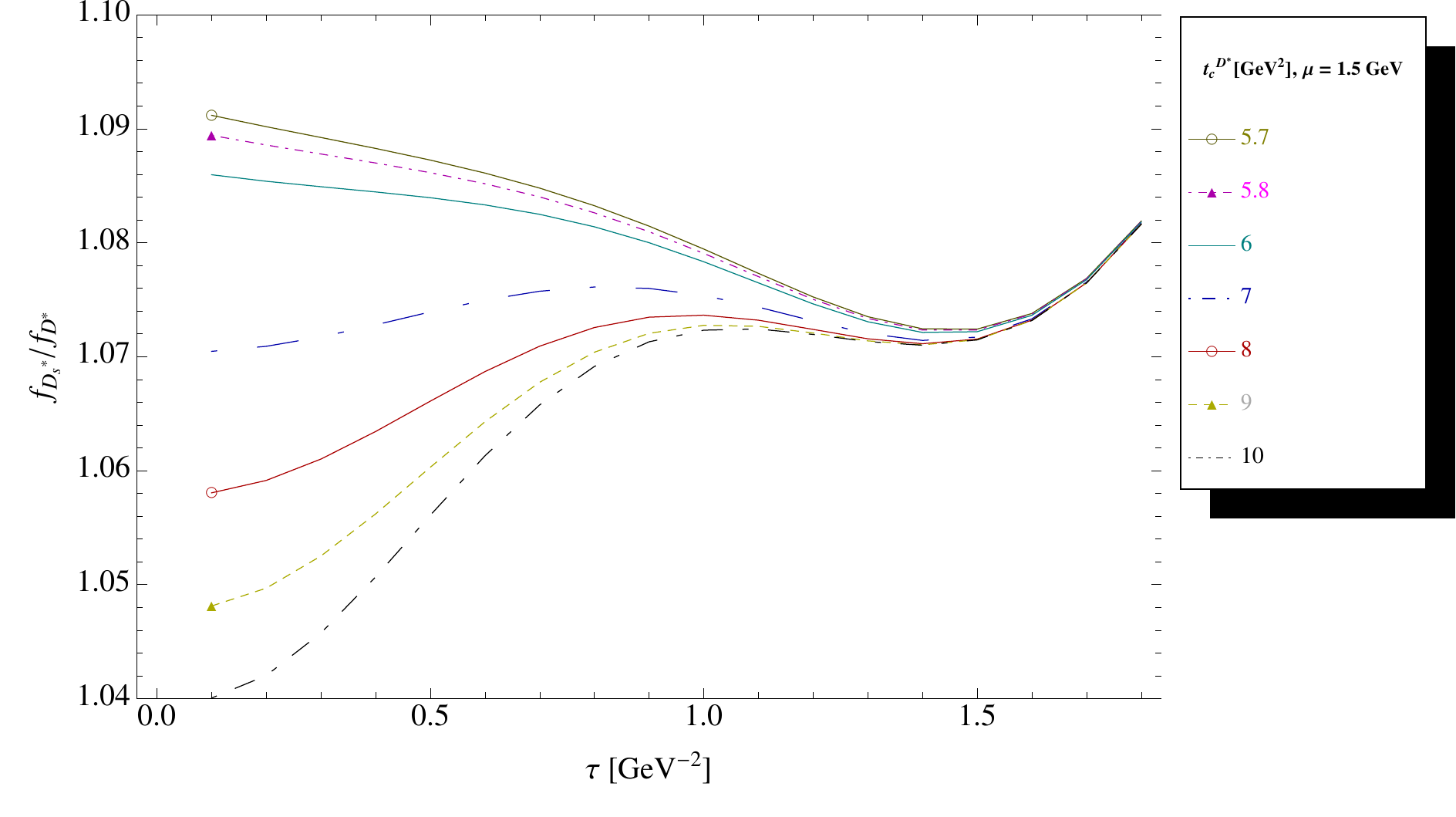}}
\caption{
\scriptsize 
 $\tau$-behaviour of $f_{D^*_s}/f_{D^*}$ for different values of $t_c$ and for $\mu=1.5$ GeV.}
\label{fig:rd*stau} 
\end{center}
\end{figure} 
\nin
%%%%%%%%%%%%%%%%%%%%%%%%%
%%%%%%%%%%%%%%%%%%%%%%%%%
\begin{figure}[hbt] 
\begin{center}
{\includegraphics[width=8cm  ]{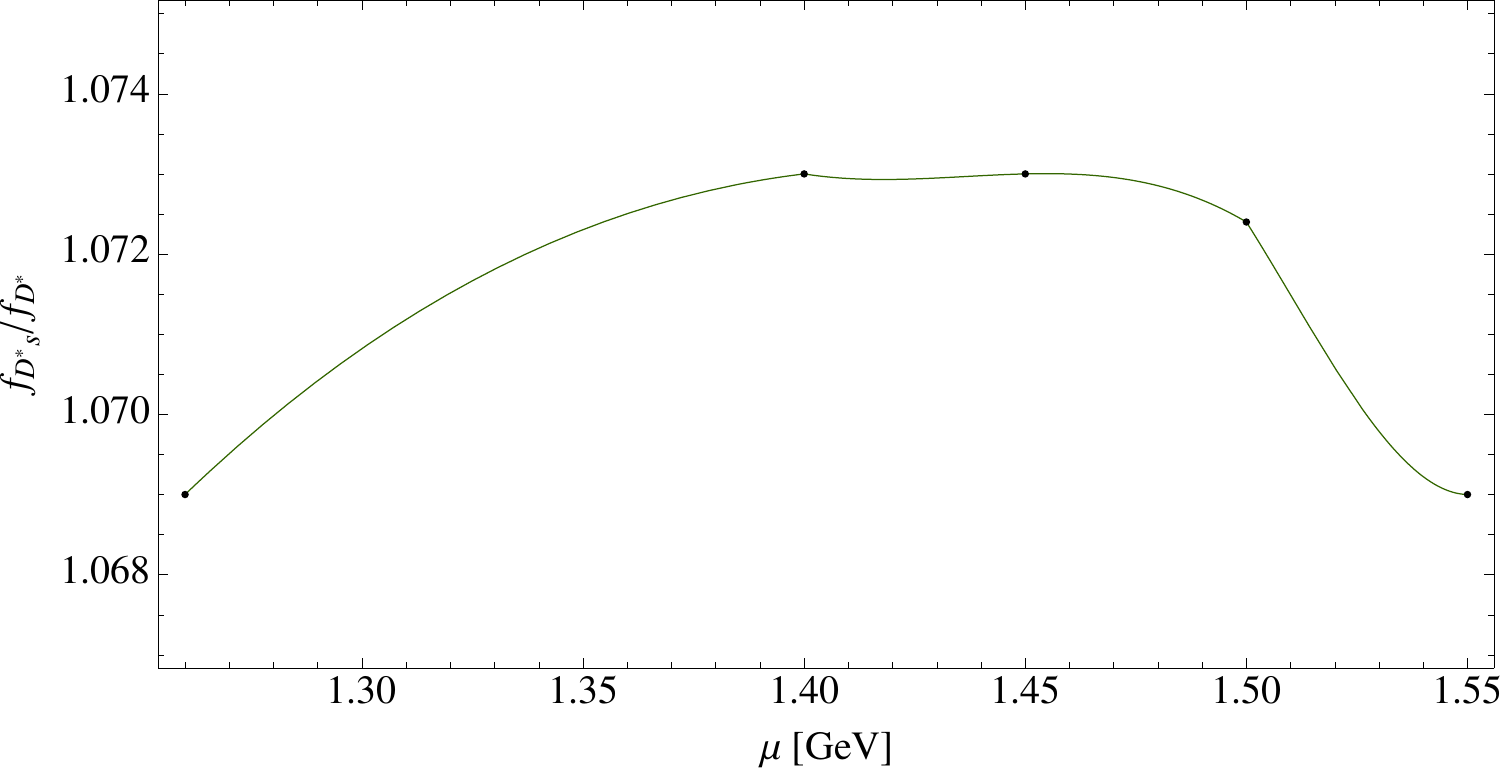}}
\caption{
\scriptsize 
 $\mu$-behaviour of $f_{D^*_s}/f_{D^*}$.}
\label{fig:rd*smu} 
\end{center}
\end{figure} 
\nin
%%%%%%%%%%%%%%%%%%%%%%%%%
where we have added the relative errors quadratically. Alternatively, we extract directly the previous ratio using the ratio of sum rules. We show the results in Fig. \ref{fig:rd*stau} versus $\tau$ and for different values of $t_c$ at $\mu=1.5$ GeV. $\tau$-stabilities occur from $\tau\simeq$ 1 to 1.5 GeV$^{-2}$. We also show in Fig \ref{fig:rd*smu} the $\mu$ behaviour of the results where a good stability in $\mu$ is observed for $\mu \simeq (1.4-1.5)$ GeV in the same way as for $f_{D^*_s}$. We deduce:
\bea
f_{D^*_s}/ f_{D^*}&=&1.073(1)_{t_c}(16)_\tau(2)_\mu(50)_{svz}\nnb\\
&=&1.073(52)~,
\lb{eq:rd*stau2}
\eea
with:
\bea
(50)_{svz}&=&(1)_{\alpha_s}(45)_{\alpha_s^3}(0)_{m_c}(2)_{\la\bar dd\ra}(16)_{\la \alpha_sG^2\ra}\nnb\\
&&(3)_{\la\bar dGd\ra}(1)_{\la g^3G^3\ra}(2)_{\la\bar dd\ra^2}\nnb\\
&&(4)_{m_s}(13)\kappa~,
\eea
where, for asymetric errors, we have taken the mean of the two extremal values. The error associated to $\tau$ take into accounts the fact that, for some values of $t_c$, the $\tau$-minima for $f_{D^*}$ and $f_{D^*_s}$ do not co\"\i ncide. Comparing the results in Eqs. (\ref{eq:rd*stau1}) and  (\ref{eq:rd*stau2}), one can clearly see the advantage of a direct extraction from the ratio of moments due to the cancellation of systematic errors in the analysis. Taking the mean of the two different results in Eqs. (\ref{eq:rd*stau1}) and  (\ref{eq:rd*stau2}),   we deduce our final estimate:
\beq
f_{D^*_s}/ f_{D^*}=1.08(6)(1)_{syst}~,
\lb{eq:fd*sd*}
\eeq
where the 1st error comes from the most precise determination and the 2nd one from the distance of the mean value to the central value of this precise determination. 
This value is in better agreement with the lattice result \cite{BECIR}: $1.16(6)$ than the one from the sum rules analysis $1.21(5)$ in \cite{PIVOV} and $1.21(7)$ in \cite{LUCHA}. The almost good agreement with the lattice result is due to the fact that both $f_{D^*}$ and $f_{D^*_s}$  are larger from the lattice than in the present paper while the ratio is less affected by
this discrepancy. The disagreement with the sum rule result of \cite{PIVOV} is due to a larger value  of $f_{D^*_s}$  in \cite{PIVOV} related to the choice of $t_c=7.4$ GeV$^2$ but to a value of of $f_{D^*}$ similar to ours because taking the same value of $t_c\simeq 5.6$ GeV$^2$.  The discrepancy with the one in \cite{LUCHA} is due to a larger value of subtraction scale $\mu=1.94$ GeV which is outside the $\mu$ stability region shown in Fig. \ref{fig:fd*smu} for $f_{D^*_s}$. In fact, one would intuitively expect that, up to small $SU(3)$ breaking corrections, the value of $\mu$ is about the same for $f_{D^*}$ and  $f_{D^*_s}$ as explicitly shown in Figs.   \ref{fig:fd*mu} and \ref{fig:fd*smu}. 
Using the value in Eq. (\ref{eq:fd*sd*}) and the prediction for $f_{D^*}$ given in Eq. (\ref{eq:fd*}), we predict:
\beq
f_{D^*_s}=270(19)~{\rm MeV}~.
\eeq
Combining the results in Eqs. (\ref{eq:fd*}) and (\ref{eq:fd*sd*}), we deduce the upper bound:
\bea
f_{D^*_s}&\leq& 287(8.6)(16)~{\rm MeV}\nnb\\
&\leq& 287(18)~{\rm MeV}~.
\eea
Future experimental measurements of  $f_{D^*}$ and $f_{D^*_s}$ though most probably quite difficult should provide a decisive selection of these existing theoretical predictions.
%%%%%%%%%%%%%%%%%%%%%%%%%%%%%%%%%%%%%%
\section{ SU(3) breaking for $f_{B^*_s}$ and $f_{B^*_s}/f_{B^*}$ }
%%%%%%%%%%%%%%%%%%%%%%%%%%%%%%%%%%%%%%
%We have seen previously that the most precise estimate of the $SU(3)$ breaking ratio of decay constants come from its direct estimate from the ratio of Laplace sum rules. 
%%%%%%%%%%%%%%%%%%%%%%%%%
%\vspace*{-0.3cm}
\begin{figure}[hbt] 
\begin{center}
{\includegraphics[width=8cm  ]{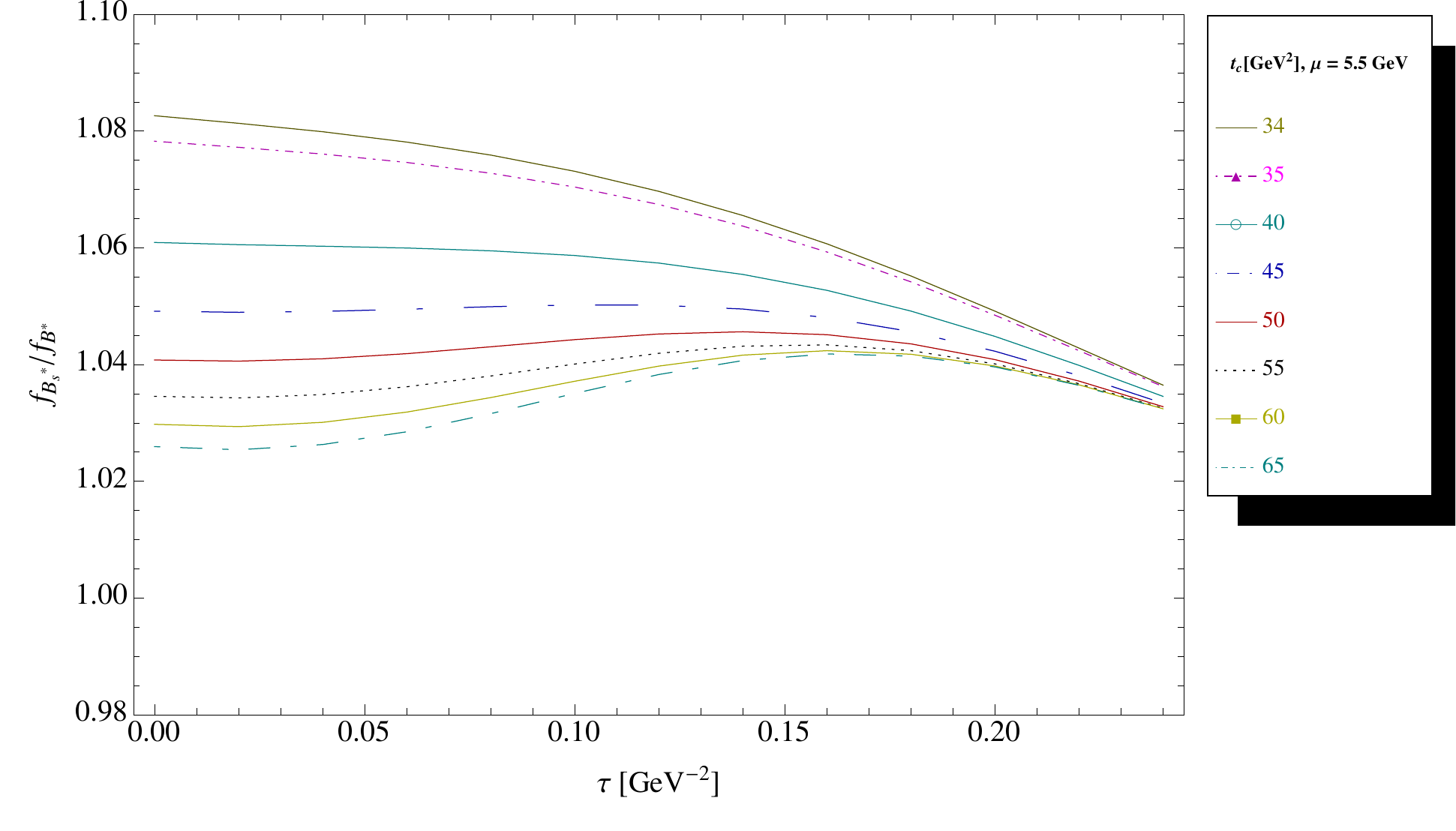}}
\caption{
\scriptsize 
$\tau$-behaviour of $f_{B^*_s}/f_{B^*_s}$ for different values of $t_c$, at a given value of the subtraction point $\mu=5$ GeV.}
\label{fig:rb*stau} 
\end{center}
\end{figure} 
\nin
%%%%%%%%%%%%%%%%%%%%%%%%%
%%%%%%%%%%%%%%%%%%%%%%%%%
%\vspace*{-0.3cm}
\begin{figure}[hbt] 
\begin{center}
{\includegraphics[width=8cm  ]{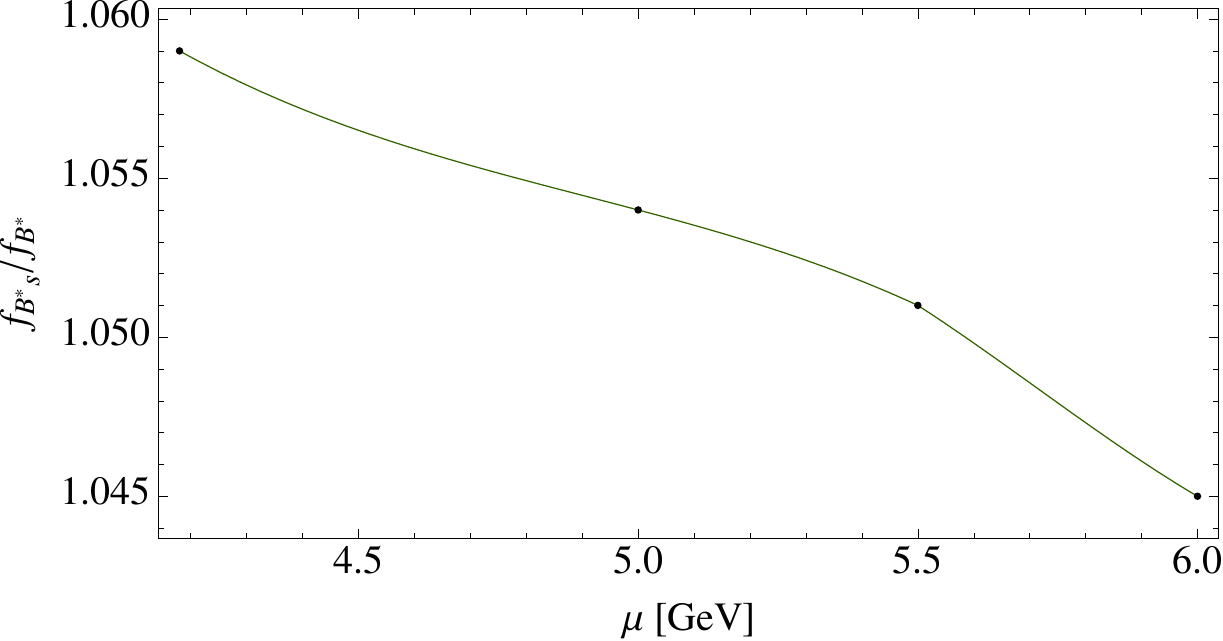}}
\caption{
\scriptsize 
 $\mu$-behaviour of $f_{B^*_s}/f_{B^*_s}$.}
\label{fig:rb*smu} 
\end{center}
\end{figure} 
\nin
%%%%%%%%%%%%%%%%%%%%%%%%%
We extend the analysis done for the $D^*_s$ to the case of the $B^*_s$-meson. 
We show, in Fig. \ref{fig:rb*stau}, the $\tau$-behaviour of the ratio $f_{B^*_s}/f_{B^*}$  at $\mu$=5 GeV and for different values of $t_c$ where the $\tau$ stability starts from $t_c=40$ GeV$^2$
while the $t_c$ one is reached for $t_c\simeq (60-65)$ GeV$^2$. Our optimal result is taken in this range of $t_c$. We study the $\mu$ behaviour in Fig. \ref{fig:rb*smu} where  an inflexion point is obtained for $\mu=(5\pm.5)$ GeV. At this point, we obtain:
\bea
f_{B^*_s}/f_{B^*}&=&1.054(8)_{t_c}(0)_\tau(6)_\mu(4.6)_{svz}\nnb\\
&=&1.054(11)~,
\lb{eq:rb*stau}
\eea
with:
\bea
(4.6)_{svz}&=&(2)_{\alpha_s}(2.5)_{\alpha_s^3}(0)_{m_b}(2)_{\la\bar dd\ra}(1.5)_{\la \alpha_sG^2\ra}\nnb\\
&&(0)_{\la\bar dGd\ra}(0)_{\la g^3G^3\ra}(0)_{\la\bar dd\ra^2}\nnb\\
&&(0)_{m_s}(2)\kappa~,
\eea
%%%%%%%%%%%%%%%%%%%%%%%%%%%%%%%%
\begin{figure}[hbt] 
\begin{center}
{\includegraphics[width=8cm  ]{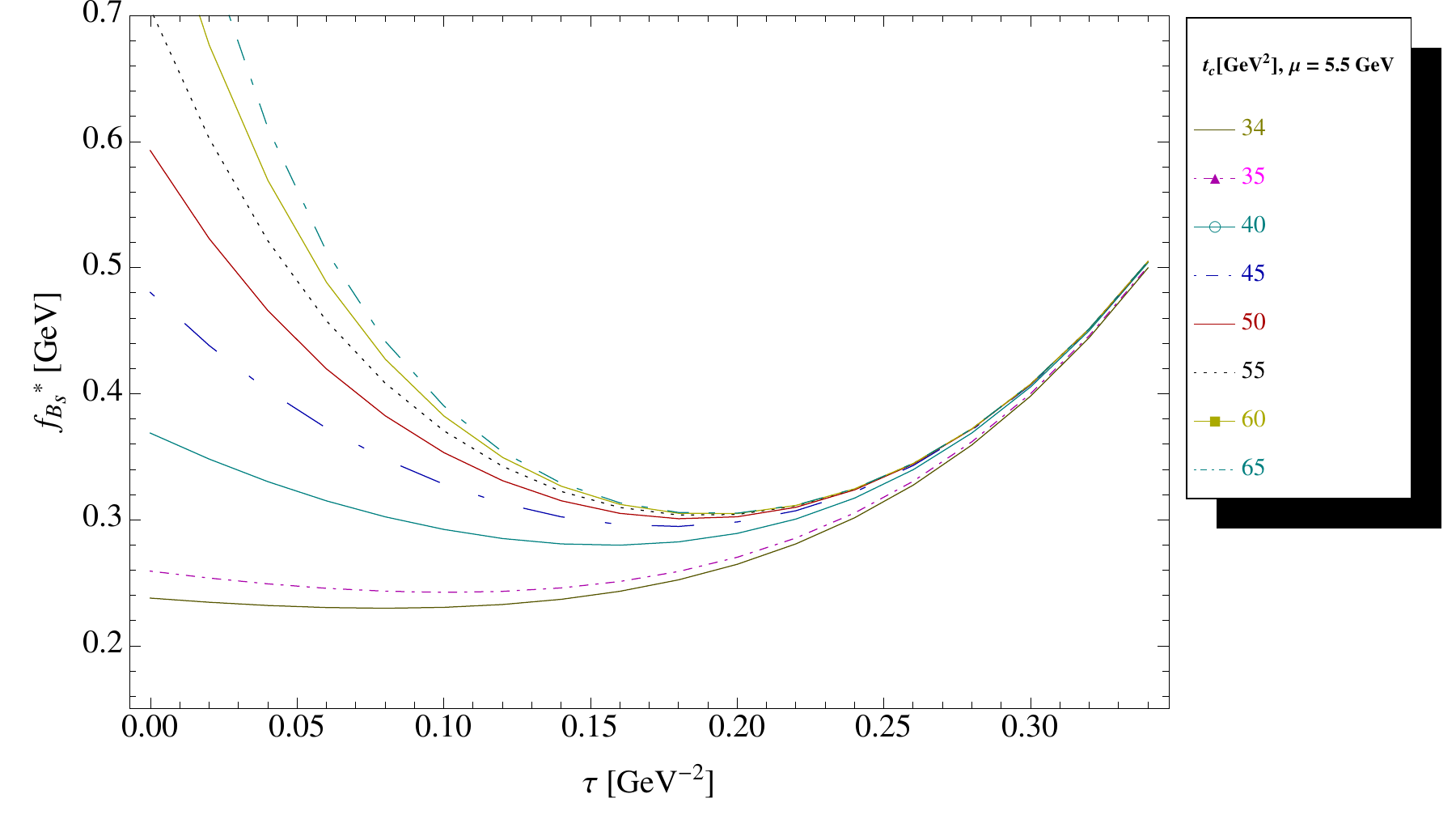}}
\caption{
\scriptsize 
 $\tau$-behaviour of $f_{B^*_s}$ for different values of $t_c$, at a given value of the subtraction point $\mu=6$ GeV.}
\label{fig:fb*stau} 
\end{center}
\end{figure} 
\nin
%%%%%%%%%%%%%%%%%%%%%%%%%
%%%%%%%%%%%%%%%%%%%%%%%%%
%\vspace*{-0.3cm}
\begin{figure}[hbt] 
\begin{center}
{\includegraphics[width=8cm  ]{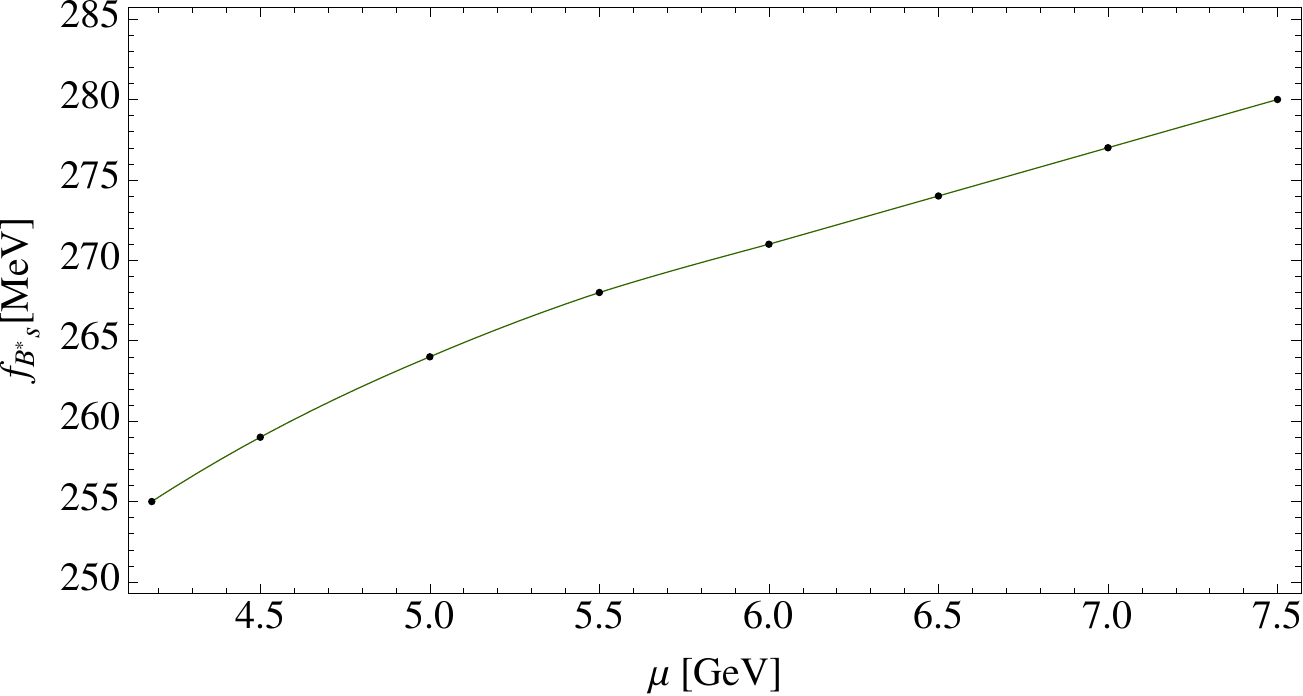}}
\caption{
\scriptsize 
$\mu$-behaviour of $f_{B^*_s}$.}
\label{fig:fb*smu} 
\end{center}
\end{figure} 
\nin
%%%%%%%%%%%%%%%%%%%%%%%%%
We show, in Fig. \ref{fig:fb*stau}, the $\tau$ behaviour of the result for $f_{B^*_s}$  at $\mu$=5.5 GeV and for different values of $t_c$. For $f_{B^*_s}$, $\tau$-stability starts from $t_c\simeq 34$ GeV$^2$ while $t_c$ stability is reached from $t_c\simeq (50-65)$ GeV$^2$.  We show in Fig. \ref{fig:fb*smu} the $\mu$ behaviour of these optimal results. One can notice a slight inflexion point  at $\mu$=6 GeV which is about the value (5.0-5.5) GeV where the ratio  $f_{B^*_s}/f_{B^*}$ has been obtained previously. At this value of $\mu$, we obtain:
\bea
f_{B^*_s}&=&271(39)_{t_c}(0)_\tau(3)_{svz}(6)_\mu~{\rm MeV}\nnb\\
&=&271(40)~{\rm MeV}~,
\eea
with:
\bea
(3)_{svz}&=&(1)_{\alpha_s}(1.5)_{\alpha_s^3}(0.5)_{m_b}(2)_{\la\bar dd\ra}(0.5)_{\la \alpha_sG^2\ra}\nnb\\
&&(0.5)_{\la\bar dGd\ra}(0)_{\la g^3G^3\ra}(0.5)_{\la\bar dd\ra^2}\nnb\\
&&(0)_{m_s}(1)\kappa~.
\eea
Combining consistently this result with the one of $f_{B^*}$ in Eq. (\ref{eq:fb*mu}) obtained within the same approach and conditions, we deduce the ratio:
\beq
f_{B^*_s}/f_{B^*}=1.13(25)~,
\lb{eq:rbs1}
\eeq
where the large error is due to the determinations of each absolute values of the decay constants. 
We take as a final value of the ratio $f_{B^*_s}/f_{B^*}$  the most precise determination in Eq. (\ref{eq:rb*stau}).
%\bea
%f_{B^*_s}/f_{B^*}&=&1.075(10)(21)_{syst}\nnb\\
%&=&1.075(23)~.
%\lb{eq:rb*s}
%\eea
Combining this result with the final value of $f_{B^*}$  in Eq. (\ref{eq:fb*}), we deduce our final estimate:
\beq
f_{B^*_s}=220(9)~{\rm MeV}~.
\lb{eq:fb*s}
\eeq
Combining again this result of the ratio with the upper bound of $f_{B^*}$ in Eq. (\ref{eq:fb*_bound}), 
we deduce the upper bound:
\beq
f_{B^*_s}\leq 311(19)~{\rm MeV}~.
\lb{eq:fb*s_bound}
\eeq
%%%%%%%%%%%%%%%%%%%%%%%%%%%%%%
\section{The decay constant $f_{B_c}$}
%%%%%%%%%%%%%%%%%%%%%%%%%%%%%%
\nin
We complete the analysis in this paper by the estimate and the bound of the decay constant $f_{B_c}$ of the $B_c(6277)$ meson $\bar bc$ bound state where the light quarks $d,s$ are replaced by the heavy quark $c$. Our analysis will be very similar to the one in \cite{BAGAN} but we shall use the running $c$ and $b$ quark masses instead of the pole masses and we shall include N2LO radiative  corrections in the analysis.

\b The dynamics of the $B_c$ is expected to be different from the $B$ and $B^*$ because, by using the heavy quark mass expansion, the heavy quark $\la \bar cc\ra$ and quark-gluon mixed $\la \bar cGc\ra$ condensates defined in Eq. (\ref{d5mix}) behave as\,\cite{BAGAN}:
\bea
\la \bar cc\ra&=&-{1\over 12\pi m_c}\la\alpha_s G^2\ra-{\la g^3 G^3\ra\over 1440\pi^2 m_c^3}~,\nnb\\
\la \bar cGc\ra&=&{m_c\over \pi}\ga\log{m_c\over \mu}\dr\la\alpha_s G^2\ra-{\la g^3 G^3\ra\over 48\pi^2 m_c}~,
\eea
in terms of the gluon condensates defined in Eqs. (\ref{d4g}) and (\ref{eq:d6g}). These behaviours are in contrast with the ones of the light quark  $\la\bar qq\ra$ and mixed quark-gluon $\la \bar qGq\ra$ condensates where $q\equiv d,s$\,\cite{SNB1,SNB2}.

\b The complete expression of the perturbative NLO spectral function has been obtained in \cite{GENERALIS} and explicitly written in \cite{BAGAN}. We add to this expression the N2LO result obtained in \cite{CHETa,CHETb} for $m_c=0$. We consider as a source of errors an estimate of the N3LO contribution by assuming a geometric growth of the PT series \cite{SZ} which mimics the phenomenological $1/q^2$ dimension-two term which parametrizes the large order terms of PT series \cite{CNZa,CNZb,ZAKa,ZAKb}. 

\b The Wilson coefficients of the non-perturbative $\la \alpha_s G^2\ra$ and $\la g^3 G^3\ra$ contributions are also given in \cite{BAGAN}. We transform the pole masses to the running masses using the previous expression in Eq. (\ref{eq:pole}). 
%%%%%%%%%%%%%%%%%%%%%%%%%%%%%%
%\subsection{Estimate of $f_{B_c}$}
%\nin
%%%%%%%%%%%%%%%%%%%%%%%%%%%

\begin{figure}[hbt] 
\begin{center}
{\includegraphics[width=9cm  ]{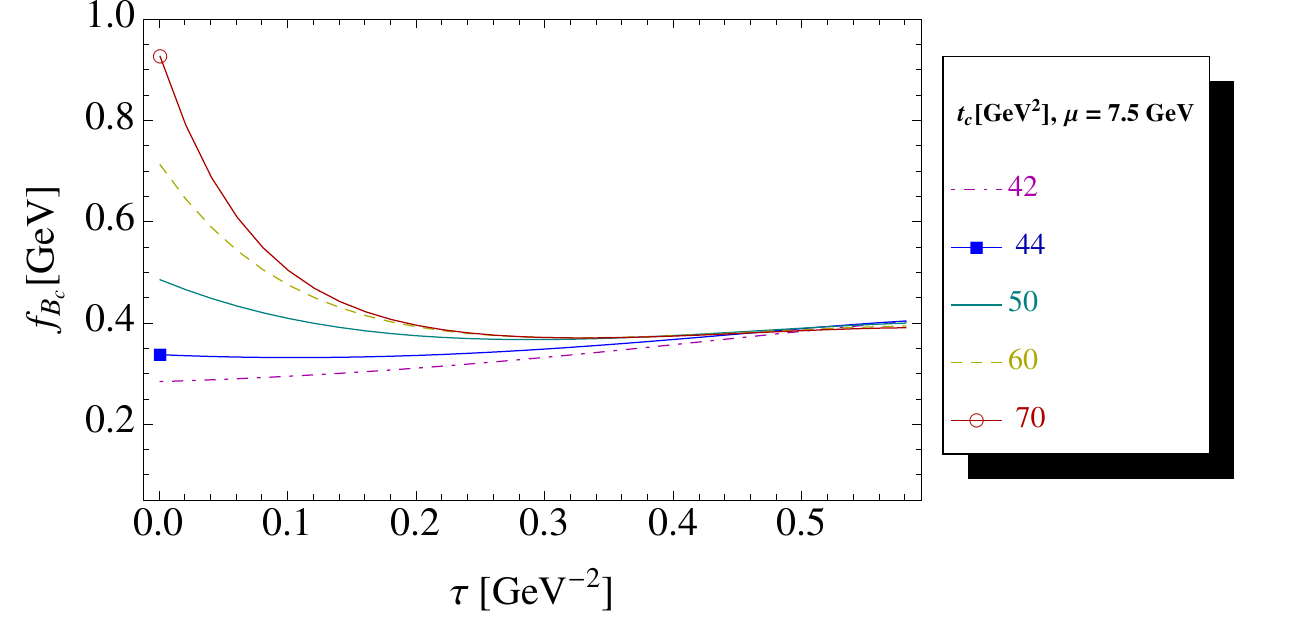}}
\caption{
\scriptsize 
 $\tau$-behaviour of $f_{B_c}$  from  ${\cal L}_{P} $ for different values of $t_c$, at a given value of the subtraction point $\mu=7.5$ GeV. }
\label{fig:fbctau} 
\end{center}
\end{figure}

\b Like in the case of $D^*$ and $B^*$ mesons, we study the corresponding (inverse) Laplace sum rule versus $\tau$ and for different values of $t_c$ which we show in Fig. \ref{fig:fbctau}. We notice that the non-perturbative contributions are small (about 1-2 MeV) indicating that the dynamics of the $B_c$ meson is dominated by the perturbative contributions. This feature might explain the success of the non-relativistic potential models for describing the $B_c$-like hadrons \cite{BAGAN}. The optimal result is obtained from $t_c=44$ GeV$^2$ (beginning of $\tau$ stability) until $t_c=(50-60)$ GeV$^2$ (beginning of $t_c$ stability). 
%This feature may indicate that one may expect that the non-relativistic approach works much better here, which has been signaled by the good prediction of the $B_c$ mass from this approach \cite{BAGAN} and confirmed by QSSR \cite{BAGAN} long time before the discovery of the $B_c$ meson. 
%%%%%%%%%%%%%%%%%%%%%%%%%
%\vspace*{-0.3cm}
\nin
%%%%%%%%%%%%%%%%%%%%%%%%
  %%%%%%%%%%%%%%%%%%%%%%%%%
\begin{figure}[hbt] 
\begin{center}
\centerline {\hspace*{-7cm} a) }%\vspace{-0.7cm}
{\includegraphics[width=7cm  ]{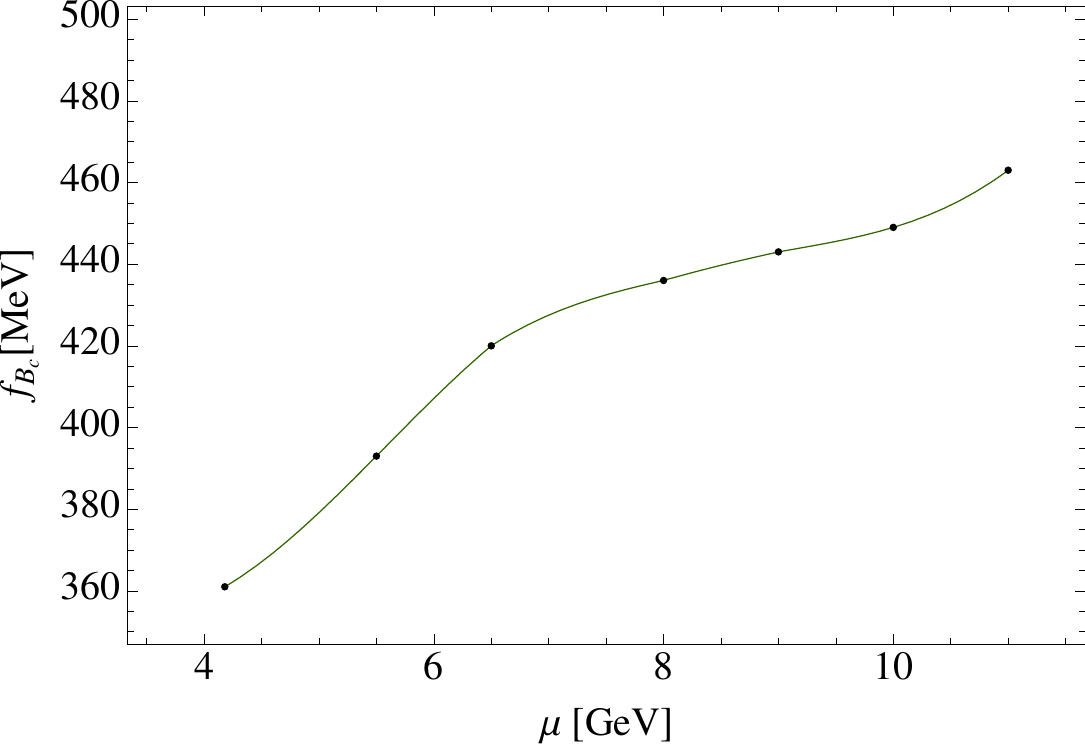}}\\
\centerline {\hspace*{-7cm} b) }%\vspace{-0.7cm}
{\includegraphics[width=7cm  ]{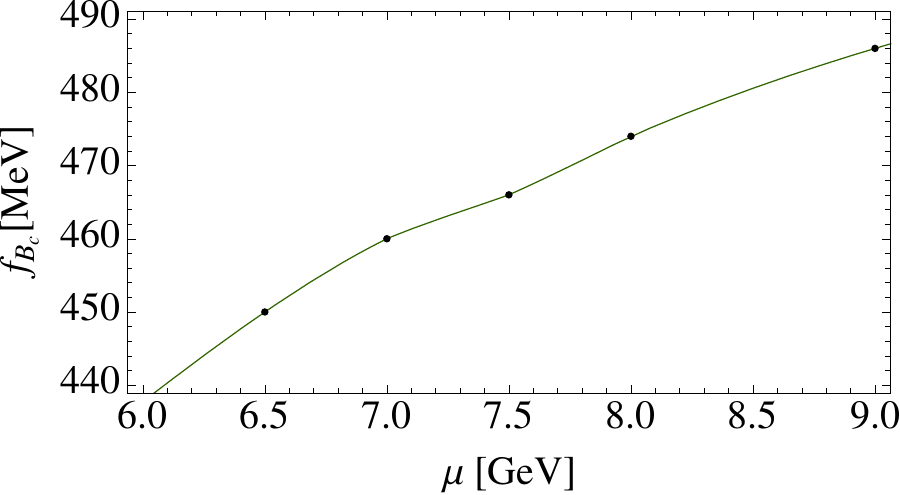}}
\caption{
\scriptsize 
Values of  $f_{B_c}$  at different values of the subtraction point $\mu$: a) estimate; b) upper bound } 
%The dashed horizontal lines are the values if one takes the errors from the best determination.}
\label{fig:fbcmu} 
\end{center}
\end{figure} 
\nin
%%%%%%%%%%%%%%%%%%%%%%%%%%%%%%%
\\
We show in Fig. \ref{fig:fbcmu} the $\mu$-behaviour of different results, where one can notice that there is an inflexion point for $\mu=(7.5\pm 0.5)$ GeV for both the estimate (Fig. \ref{fig:fbcmu}a) and the upper bound (Fig. \ref{fig:fbcmu}b).  At these optimal points, we deduce:
\bea
f_{B_c}&=&436(38)_{t_c}(2)_{\alpha_s}(2)_{\alpha_s^3}(7)_{m_c}(6)_\mu~{\rm MeV}\nnb\\
&=&436(40)~{\rm MeV}~.
\label{eq:fbc}
\eea
and 
\bea
f_{B_c}&\leq& 466(9)_{\alpha_s}(2)_{\alpha_s^3}(12)_{m_c}(8)_\mu~{\rm MeV}\nnb\\
&\leq&466(16)~{\rm MeV}~.
\label{eq:fbc_bound}
\eea
We may consider the previous results as  a confirmation and improvement of the earlier ones  obtained in \cite{SNB2,SNFB88,BAGAN,CHABABa,CHABABb}. Having in mind that, long time before the experimental discovery of $B_c$, the correct prediction of $M_{B_c}$ from QSSR has been given in \cite{BAGAN} together with some potential model predictions,  which was not the case of some early lattice results, the agreement of our results  in Eq. (\ref{eq:fbc})  with the recent lattice value $f_{B_c}=(427\pm 6)$ MeV  in \cite{LATTFBC} (and to a lesser extent with the large range of potential model predictions $f_{B_c}=(503\pm 171)$ MeV reviewed in \cite{BAGAN}) can be considered as a strong support of our results and  may question the validity of a recent estimate $f_{B_c}=(528\pm 19)$ MeV from some variants of FESR \cite{DOM} evaluated at a given $\mu$. This discrepancy might be due to  the induced systematic uncertainties not properly accounted for when requiring that the {\it  input Legendre polynomial kernel integral} from the continuum threshold $t_c$ to the arbitrary FESR cut-off $s_0$ vanishes in this channel where the results increase with the degree $n$ of the polynomial and with $s_0$ (Table 1 and Fig. 1 of Ref. \cite{DOM}). \\
Our previous estimate in Eq. (\ref{eq:fbc}) and the upper bound in Eq. (\ref{eq:fbc_bound}) together with the recent lattice result will restrict the wide range of $f_{B_c}$ values given in the current literature and may be used for extracting the CKM angle $V_{cb}$ from the predicted leptonic width:
\beq
\Gamma(B_c \to \tau \nu_\tau) = (3.9 \pm 0.7)\ga{{\rm V_{cb}}\over 0.037}\dr^2\ga{f_{B_c}[{\rm MeV}]\over 436}\dr^2,
\eeq
 in units of $10^{10}~{\rm s}^{-1}$. Our predictions for $f_{B_c}$ are compared in Table \ref{tab:res} with the ones in \cite{BAGAN} using an analogous approach but with a pole masses for the $c$ and $b$ quarks and with some other recent determinations. 

  %%%%%%%%%%%%%%%%%%%%%%%%%%%%%%%%%%%%%%%%%%%
{\scriptsize
\begin{table}[hbt]
 \tbl{    
 Estimates and upper bounds of the decay constants and comparisons with recent sum rules and lattice results.}
    {\small
%\begin{tabular}{lccccc}
\begin{tabular}{@{}lccccc@{}} \toprule
&\\
\hline
\hline
%$f_M$&$t_c$ [GeV]$^2$&$\mu$ [GeV]&Value [MeV]&Sources &Ref. \\
&$t_c$ [GeV]$^2$&$\mu$ [GeV]&Sources &Refs. \\
\cline{2-5}
%\hline
\boldmath$f_{D^*}$ [MeV] \\
250(11)&$5.6-10.5$&1.5&Eq.(\ref{eq:fd*})&This work\\
$242^{+20}_{-12}$&6.2&1.5&SR&\cite{PIVOV}\\
252(22)&5.52&1.84&SR&\cite{LUCHA}\\
278(16)&&&Latt.&\cite{BECIR}\\
$\leq 266(8)$&$\infty$&$1.5-2.0$&Eqs.(\ref{eq:fdbound1},\ref{eq:fdbound2})&This work\\
$\leq 297$&$\infty$&$1.5$&SR&\cite{PIVOV}\\
%\\
%\multicolumn{2}{c}{${f_{D^*}/f_D}$} &&&&\\
\boldmath{$f_{D^*}/f_{D}$} &&&&\\
1.218(36)&$5.6-10.5$&1.5&Eq. (\ref{eq:fdd*})&This work\\
$1.20^{+0.10}_{-0.07}$&6.2&1.5&SR&\cite{PIVOV}\\
1.221(80)&5.52&1.84&SR&\cite{LUCHA}\\
\boldmath$f_{B^*}$ [MeV] \\
209(8)&$34-60$&$5-5.5$&Eq.(\ref{eq:fb*})&This work\\
$210^{+10}_{-12}$&34.1&3&SR&\cite{PIVOV}\\
$\leq 295(18)$&$\infty$&$1.5-2.0$&Eqs.(\ref{eq:fb*_bound})&This work\\
$\leq 261$&$\infty$&$3$&SR&\cite{PIVOV}\\
%\\
%\multicolumn{2}{c}{${f_{D^*}/f_D}$} &&&&\\
%%%%%%%%%%%%%%%%%%%%%%%%%%%%%%
\boldmath{$f_{B^*}/f_{B}$} &&&&\\
1.016(16)&$34-60$&$5-5.5$&Eq. (\ref{eq:rb*mu})&This work\\
$1.02^{+0.07}_{-0.03}$&$34-36$&3&SR&\cite{PIVOV}\\
%%%%%%%%%%%%%%%%%%%%%%%%%%%%%%
\boldmath{$f_{B_c}$} [MeV]&&&&\\
436(40)&$44-60$&$7.5\pm 0.5$&Eq. (\ref{eq:fbc})&This work\\
$\leq 466(16)$&$\infty$&$-$&Eq. (\ref{eq:fbc_bound})&--\\
383(27)&&$\tau^{-1/2}$&SR Pole &\cite{BAGAN}\\
503(171)&&&Pot. Mod.&\cite{BAGAN}\\
427(6)&&&Latt. &\cite{LATTFBC}\\
528(19)&50.6&&SR &\cite{DOM}\\
\\

\hline
\hline
\end{tabular}
}
\label{tab:res}
\end{table}
} 
%\end{document}
%%%%%%%%%%%%%%%%%%%%%%%%%
  %%%%%%%%%%%%%%%%%%%%%%%%%%%%%%%%%%%%%%%%%%%
{\scriptsize
\begin{table}[hbt]
 \tbl{    
$SU(3)$ breaking effects on the  estimates and on the upper bounds of the decay constants and comparisons with recent sum rules and lattice results.} 
    {\small
    \begin{tabular}{@{}lccccc@{}} \toprule
%\begin{tabular}{lccccc}
&\\
\hline
\hline
%$f_M$&$t_c$ [GeV]$^2$&$\mu$ [GeV]&Value [MeV]&Sources &Ref. \\
&$t_c$ [GeV]$^2$&$\mu$ [GeV]&Sources &Refs. \\
\cline{2-5}
%\hline
\boldmath$f_{D^*_s}$ [MeV] \\
270(19)&$5.6-10.5$&1.5&Eq.(\ref{eq:fd*})&This work\\
$293^{+19}_{-14}$&7.4&1.5&SR&\cite{PIVOV}\\
306(27)&5.52&1.94&SR&\cite{LUCHA}\\
311(9)&&&Latt.&\cite{BECIR}\\
$\leq 287(18)$&$\infty$&$1.5-2.0$&Eqs.(\ref{eq:fdbound1},\ref{eq:fdbound2})&This work\\
$\leq 347$&$\infty$&$1.5$&SR&\cite{PIVOV}\\
%\\
%\multicolumn{2}{c}{${f_{D^*}/f_D}$} &&&&\\
\boldmath{$f_{D^*_s}/f_{D^*}$} &&&&\\
1.08(6)&$5.6-10.5$&1.5&Eq. (\ref{eq:fdd*})&This work\\
$1.21^{+0.06}_{-0.04}$&$6.2-7.4$&1.5&SR&\cite{PIVOV}\\
1.21(6)&5.52&1.84&SR&\cite{LUCHA}\\
1.16(6)&&&Latt.&\cite{BECIR}\\
\boldmath$f_{B^*_s}$ [MeV] \\
220(9)&$34-60$&$6\pm 0.5$&Eq.(\ref{eq:fb*s})&This work\\
$210^{+10}_{-12}$&34.1&3&SR&\cite{PIVOV}\\
$\leq 317(17)$&$\infty$&$6\pm 0.5$&Eq.(\ref{eq:fb*s_bound})&This work\\
$\leq 296$&$\infty$&$3$&SR&\cite{PIVOV}\\
%\\
%\multicolumn{2}{c}{${f_{D^*}/f_D}$} &&&&\\
\boldmath{$f_{B^*_s}/f_{B^*}$} &&&&\\
1.054(11)&$34-60$&$5\pm 0.5$&Eq. (\ref{eq:rb*stau})&This work\\
$1.20(4)$&$34-36$&3&SR&\cite{PIVOV}\\
\\

\hline
\hline
\end{tabular}
}
\label{tab:su3}
\end{table}
} 
%%%%%%%%%%%%%%%%%%%%%%%%%
%%%%%%%%%%%%%%%%%%%%%%%%%%%%
\section*{Summary and conclusions}
%%%%%%%%%%%%%%%%%%%%%%%%%%%%%
\nin
Our main results are summarized in Tables \ref{tab:res} and \ref{tab:su3} where a comparison with some other recent sum rules and lattice results is done.

\b We have re-estimated $f_{D^*}$ and $f_{B^*}$ directly from the Laplace sum rule of vector current and indirectly by combining our previous results for $f_D$ and $f_B$ \cite{SNFB12a,SNFB12b} with suitable ratios of Laplace sum rules for $f_{D^*}/f_D$ and $f_{B^*}/f_B$  known to N2LO of PT, including complete non-perturbative contributions of dimension 6 and an estimate of the N3LO PT-term where for the latter a geometric growth of the PT coefficients has been assumed. Our results  given in Eqs. (\ref{eq:fdd*}), (\ref{eq:fd*}), (\ref{eq:rb*mu}) and (\ref{eq:fb*})   agree and improve our earlier determinations in \cite{SNFB88,SNB2} and agree with some recent sum rule estimates\,\cite{PIVOV,LUCHA} obtained at a particular value of the continuum threshold $t_c$ where the positive large error in Ref. \cite{PIVOV} is mainly due to the arbitrary chosen large range of the subtraction region far outside the stability region. The accuracy reached here is relatively similar to the one obtained in \cite{SNFB12a,SNFB12b} for determining $f_{D,B}$ and $f_{D_s,B_s}$ and in \cite{SNH10a,SNH10b,SNH10c} for $m_{b,c}$ using similar approaches. 

\b These results indicate a good realization of heavy quark symmetry for the $B$ and $B^*$ mesons ($f_B\simeq f_{B^*}$) as expected from HQET \cite{NEUBERT2} but signal  large charm quark mass and radiative QCD corrections for the $D$ and $D^*$ mesons.

\b Our value of the $SU(3)$ breaking ratio of decay constants $f_{D^*_s}/f_{D^*}$  in Eq. (\ref{eq:fd*sd*})  disagrees with the larger value given by \cite{PIVOV} and \cite{LUCHA} correlated to a larger value of $f_{D^*}$ obtained there but is in a better agreement with the lattice result  \cite{BECIR} though the absolute values of the decay constants from lattice are individually larger. The same feature is observed for our value of $f_{B^*_s}/f_{B^*}$ in Eq. (\ref{eq:rb*stau}) when compared with the result of \cite{PIVOV}.
We expect that future experimental measurements of these couplings may select among these theoretical predictions.

\b Using the positivity of the spectral functions, we have also derived in Eqs. (\ref{eq:fdbound1}), (\ref{eq:fdbound2}), (\ref{eq:fb*_bound}) and (\ref{eq:fbc_bound}) upper bounds for $f_{D^*},~f_{B^*}$ and $f_{B_c}$. Combining these upper bounds with our estimate of the ratios $f_{D^*_s}/f_{D^*}$ and $f_{B^*_s}/f_{B^*}$, we have also derived, in Eqs.(\ref{eq:fdbound1}), (\ref{eq:fdbound2}) and  (\ref{eq:fb*s_bound}), upper bounds on $f_{D^*_s}$ and  $f_{B^*_s}$. We notice that the recent lattice result for $f_{D^*}$ \cite{BECIR} is at the borderline of the previous upper bound.

\b For completing our present study of open bottom states and motivated by the wide range of predictions in the existing literature, we have re-estimated $f_{B_c}$ by working with NLO spectral function with massive quark.  We have  added the N2LO terms with massless quark  and an estimate of the N3LO contribution based on the geometric growth of the PT coefficients. The estimate in Eq. (\ref{eq:fbc}) and the upper bound in Eq. (\ref{eq:fbc_bound}) may be considered as improvements of the ones obtained earlier from QCD spectral sum rules in  \cite{SNB2,SNFB88,BAGAN,CHABABa,CHABABb}. Comparing $f_{B_c}$ and $f_B$ which differs by about a factor two, we conclude a large $SU(4)$ breaking of the leptonic decay constant. 

\begin{figure}[hbt] 
\begin{center}
{\includegraphics[width=8cm  ]{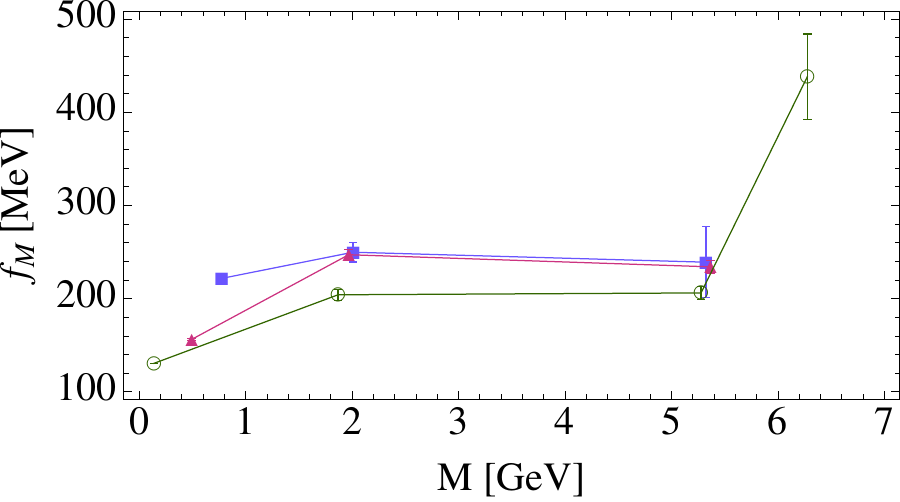}}
\caption{
\scriptsize 
 Behaviour of the meson decay constants versus the meson masses: 
 open circle (green):  $f_\pi,~f_D,~f_B$ and $f_{B_c}$ ; triangle (red): $SU(3)$ breaking: $f_K,~f_{D_{s}}$ and  $f_{B_{s}}$; boxes (blue): $f_\rho,~f_{D^*}$ and $f_{B^*}$.
 }
\label{fig:fp} 
\end{center}
\end{figure}

\b It is informative to show the behaviour of the pseudoscalar and vector meson decay constants versus the corresponding meson masses in Fig. \ref{fig:fp}. The open circles correspond to  $f_\pi,~f_D,~f_B$ and $f_{B_c}$. The triangles correspond to the one with $SU(3)$ breaking: $f_K,~f_{D_s}$ and $f_{B_s}$. The boxes correspond to $f_\rho,~f_{D^*}$ and $f_{B^*}$. $SU(3)$ breaking in the vector channels are quite small. The values of $f_{D_{(s)}}$ and  $f_{B_{(s)}}$ come from \cite{SNFB12a,SNFB12b,SNFB13} while $f_{\pi,K}$ comes from \cite{ROSNERa,ROSNERb}. We use $f_\rho=(221.6\pm1.0)$ MeV extracted from its electronic width compiled by \cite{PDG}. One can remark similar $M$ behaviours of these different couplings where the results for the $D^{(*)}_{(s)}$ and $B^{(*)}_{(s)}$ mesons do not satisfy the $1/\sqrt{M_Q}$ HQET relation.
%%%%%%%%%%%%%%%%%%%%%%%%%
%\vspace*{0.1cm}
\nin
%%%%%%%%%%%%%%%%%%%%%%%%
\vfill\eject
%\section*{References}
%\bibliography{apssamp}
%%%%%%%%%%%%%%%%%%%%%%%%%%%%% 
%\begin{thebibliography}{99}
%\end{thebibliography}

\end{document}